%% file: QDCfit-NIM-R3.tex
\documentclass[a4paper,english]{elsarticle}
\usepackage[T1]{fontenc}
\usepackage[latin9]{inputenc}
\usepackage{varioref}
\usepackage{units}
\usepackage{amsmath}
\usepackage{amssymb}
\usepackage{graphicx}
\usepackage{esint}
\usepackage{caption}
\usepackage{subcaption}

\makeatletter

\pdfpageheight\paperheight
\pdfpagewidth\paperwidth

\journal{Elsevier}
\usepackage{upgreek}
\usepackage[bookmarks,bookmarksopen,bookmarksdepth=6]{hyperref}
\usepackage{cases}
\usepackage{url}
\usepackage{lineno}
\usepackage{a4wide}
\makeatother

\usepackage{babel}
\begin{document}

\title{On the characterisation of SiPMs from pulse-height spectra}

\author[]{V.~Chmill$^{a, b}$}
\author[]{E.~Garutti$^{a}$}
\author[]{R.~Klanner$^{a,}$ \corref{cor1}}
\author[]{M.~Nitschke$^{a}$}
\author[]{and J.~Schwandt$^{a}$}

\cortext[cor1]{Corresponding author, Email: Robert.Klanner@desy.de,
  Tel. +49 40 8998 2558.}
\address{$^a$ Institute for Experimental Physics, University of Hamburg,
 \\Luruper Chaussee 149, D\,22761, Hamburg, Germany.}
\address{$^b$ Samara State Aerospace University,
 \\Moskovskoe Shosse 34, Samara 443086, Russian Federation.}

%\pdfbookmark[5]{Abstract}{abstract}}
% Analyses and plots can be found in folder Junk\detectorWork\SiPM\KETEK\N-Irradiation\Analysis\MP15\
\begin{abstract}

  Methods are developed, which use the pulse-height spectra of SiPMs measured in the dark and illuminated by pulsed light, to determine the pulse shape, the dark-count rate, the gain, the average number of photons initiating a Geiger discharge, the probabilities for prompt cross-talk and after-pulses, as well as the electronics noise and  the gain fluctuations between and in pixels.
  The entire pulse-height spectra, including the background regions in-between the peaks corresponding to different number of Geiger discharges, are described by single functions.
  As a demonstration, the model is used to characterise a KETEK SiPM with 4384 pixels of $15\,\upmu $m$ \times 15\,\upmu $m area for voltages between 2.5 and 8\,V above the breakdown voltage.
%   at $20\,^\circ$C.
  The results are compared to other methods of characterising  SiPMs.
  Finally, examples are given, how  the complete description of the pulse-eight spectra can be used to optimise the operating voltage of SiPMs, and a method for an in-situ calibration and monitoring of SiPMs, suited for large-scale applications, is proposed.

\end{abstract}


\maketitle
 \tableofcontents
% \newpage
 \pagenumbering{arabic}

\newpage

\section{Introduction}
 \label{sect:Introduction}
 Silicon photomultipliers (SiPMs), pixel arrays of avalanche photodiodes operated above the breakdown voltage, are becoming the photon detectors of choice for many applications\,\cite{Buzhan:2003,Haba:2008,Renker:2008}.
 They are robust, have a high photon-detection efficiency, achieve single-photon detection and resolution, operate at modest voltages, are not affected by magnetic fields, and are relatively inexpensive.
 Disadvantages are their high dark-count rate at room temperature, which rapidly increases with radiation damage, and the excess noise due to inter-pixel cross-talk and after-pulses.

 Various methods have been developed to determine the main SiPM-performance parameters\,\cite{Eckert:2010,Piemonte:2012,Biland:2014,Arosio:2014,Xu:2014,Xu:Thesis}:
 the photon-detection efficiency (\textit{PDE}), the gain (\textit{Gain}), the break-down voltage ($V_{bd}$), the dark-count rate (\textit{DCR}), and the correlated noise ($CN$).
 Typically, pulse-height spectra recorded in the dark are used to determine \textit{DCR} and \textit{CN}:
 $DCR = f_{0.5} / t_{gate}$, where $ f_{0.5} $ is the fraction of events above one half of the mean pulse-height ($PH$) of a single Geiger discharge and $t_{gate}$ the gate width, and
 $CN = f_{1.5}/f_{0.5}$, where $ f_{1.5} $ is the fraction of events above 1.5 times the $PH$ of a single Geiger discharge.
 From the $PH$\,spectra recorded with light from a pulsed LED or laser, the mean number of photons initiating a Geiger discharge, $N_\gamma $, which is proportional to the \textit{PDE}, the \textit{Gain}, the electronics noise, $\sigma_0$, and the gain differences between the different pixels and within one pixel, $\sigma_1$, are obtained.
 The \textit{} spectra are analysed by fitting multiple Gauss functions to the individual peaks, ignoring the background events in-between the peaks.
 The problem of the method is that the fit regions have to be selected and that it is difficult to evaluate the influence of the ignored background  on the results.
 In order to avoid the selection of event regions, methods based on Fourier transforms are also used for the \textit{Gain} determination.
 In Refs.\,\cite{Piemonte:2012} and \cite{Biland:2014} long transient are recorded in the dark and under continuous low-light illumination.
 Pulses are identified and analysed by software, and a complete characterisation, including signal shape, gain, primary dark-count rate, prompt and delayed cross-talk probability, after-pulse probability and excess noise factor, is achieved.

 In this paper, models are developed, which describe the entire \textit{PH}\,spectra:
  The model for the pulsed-light data takes into account the statistics of the photons initiating Geiger discharges, the statistics of prompt cross-talk and the pulse-height distribution and statistics of after-pulses.
 The model for the data without light takes into account the random arrival times of dark pulses and the effects of prompt cross-talk.
  By fitting the models to the measurements, a complete SiPM characterisation is achieved.

 Finally, the results of the fits are used to determine the voltage for the optimal resolution of the number of photons ($N_\gamma $), and to develop a method, which allows to determine \textit{Gain} and $N_\gamma $ in situations in which the peaks corresponding to different numbers of photoelectrons cannot be resolved.
 The method does not require fits but only uses the SiPM excess-noise factor, and the first and second moments of the measured \textit{PH} distributions.

 \section{Sensors investigated and measurements}
  \label{sect:Sensors}

\begin{figure}[!ht]
   \centering
%   \begin{subfigure}[a]{0.38\textwidth}
    \includegraphics[width=0.4\textwidth]{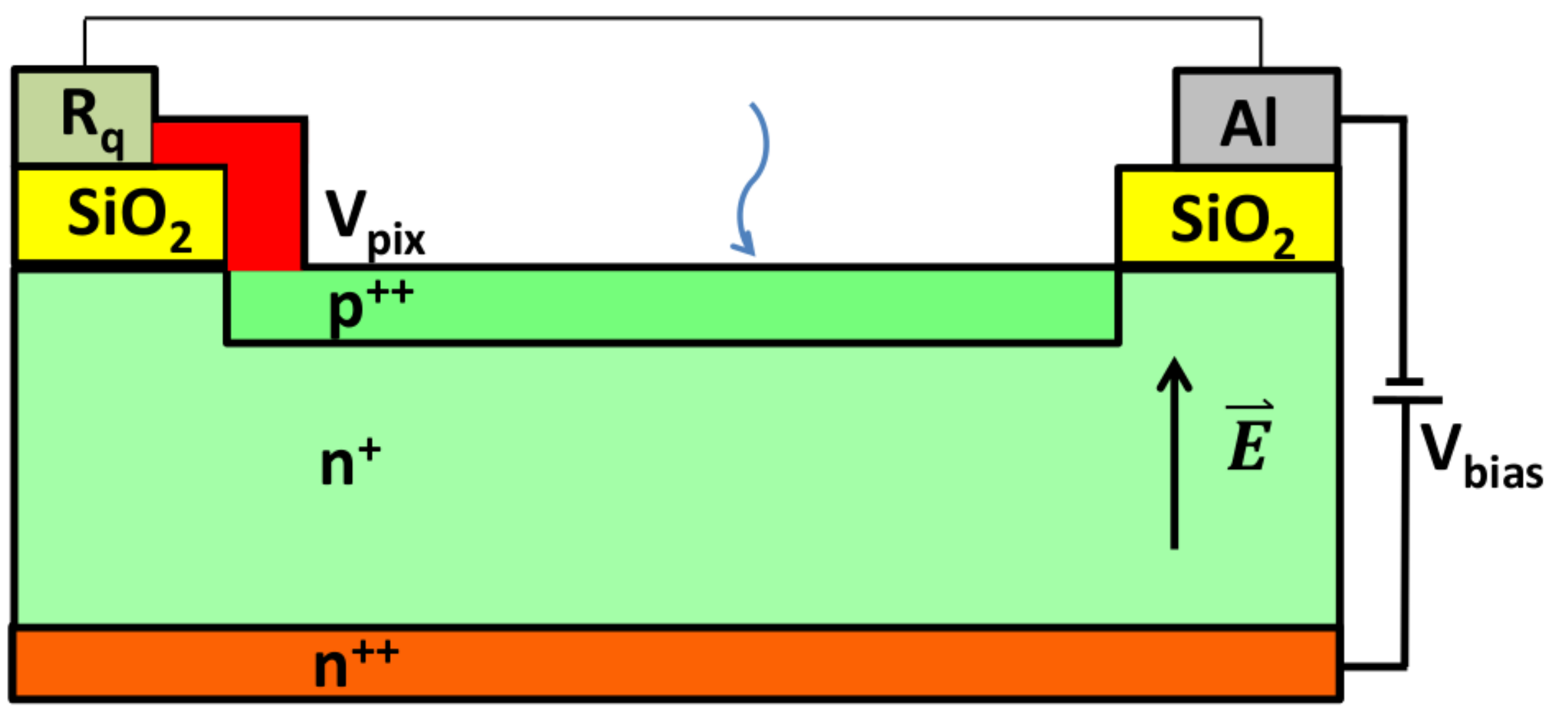}
%    \caption{ }
%   \end{subfigure}%
%    ~
%   \begin{subfigure}[a]{0.2\textwidth}
%    \includegraphics[width=\textwidth]{Fig1b.pdf}
%    \includegraphics[width=2.5cm]{Fig1b.pdf}
%    \caption{ }
%   \end{subfigure}%
   \caption{ Schematic cross section of a single pixel of the KETEK SiPM. }
  \label{fig:SiPM}
 \end{figure}

 The SiPMs investigated were fabricated by KETEK\,\cite{KETEK}.
 Their number of pixels is $N_{pix} = 4384$, and their pitch is $15\,\upmu $m$\times 15\,\upmu$m.
 Fig.\,\ref{fig:SiPM} shows a schematic cross section of a single SiPM pixel, which has a
 high-field charge-amplification region of about $1\,\upmu$m depth.
 More information on the SiPMs can be found in Refs.\,\cite{Chmill:2016,Nitschke:2016}.

 The measurements were made in a temperature-controlled light-tight box.
 As light source, a pulsed LED with a wavelength of 470\,nm and a pulse width of about 2\,ns has been used.
 The LED was housed outside of the light-tight box, and an optical fiber guided the light to the SiPM.
 The SiPM was read out via a Philips Scientific Amplifier 6954\,\cite{Philips} by a CAEN QDC V965\,\cite{QDC}.
 The read-out was AC\,coupled with a time constant $\tau_{\mathrm{AC}} \approx 5\,\upmu $s.
 The LED was triggered by a HP\,8110A Pulse Pattern Generator\,\cite{HP} by a pulse of 5\,ns full width, and a pulse generator from Stanford Research Enterprise\,\cite{Stanford} provided the 100\,ns wide gate for the QDC.
% The LED was triggered by a DG645 pulse generator from Stanford Research Enterprise\,\cite{Stanford}, which also provided the 100\,ns wide gate for the QDC.

 The following measurements were performed at 20$^\circ $C for voltages between 29.5 and 35\,V in 0.5\,V steps, which is above the breakdown voltage $V_{bd} \approx 27\,$V:
 \begin{enumerate}
   \item Delay curve,
% in order to determine the pulse shape, the AC-coupling time constant, and the effective gate width and gate delay,
   \item Pulse-height (\textit{PH}) spectra in the dark,
% in order to determine  \textit{DCR} and \textit{CN}
   \item $PH$\,spectra for low light intensity, corresponding to an average between 0.8 and 1.6 Geiger discharges in the voltage range of the measurements,
   \item $PH$\,spectra for high light intensity, $\approx 16$\,times higher than for low light.
 \end{enumerate}

 For every \textit{PH} spectrum $5 \times 10^5$ events were recorded.
 The absolute light yield of the LED is not known, however for a given voltage scan it was stable within <\,1\,\%.

 \section{Data analysis and results}
  \label{sect:Analysis}

 \subsection{Delay curve}
  \label{sect:Delay}

  The delay curves allow to determine the timing of the gate relative to the SiPM pulse, the effective gate width for the pulse-height measurement, $t_{gate}^{eff}$, the decay time constant of the SiPM signal including the effects of after-pulsing, $\tau$, and the time constant of the AC\,coupling of the readout, $\tau_{AC}$, which is typically used when reading out SiPMs.
%  The latter information is required for understanding the change in the pedestal distribution after irradiation.
%  We call $\langle PH(t) \rangle$ the average of the pulse-height spectrum recorded by the QDC, and $t$ the time between the start of the SiPM pulse and the start of the integration by the gate.
%  We assume that the SiPM-current pulse is given by $I(t') = (Q_0 /\tau) \cdot e^{-t'/\tau}$ for $t' > 0$ and $I(t') = 0$ for $t' \leq 0$.
 We call $\langle PH_{meas}(t) \rangle$ the average of the pulse-height spectrum recorded by the QDC for a measurement at a given $t$, which is the time between the start of the SiPM pulse and the start of the integration by the gate.
 For modeling the delay curve, we consider a SiPM-current pulse $I(t') = (Q_0 /\tau) \cdot e^{-t'/\tau}$ for $t' > 0$ and $I(t') = 0$ for $t' \leq 0$.
 The integrals over the gate of a pulse starting at time $t$ before the start of the integration, $t<0$, and after the start of the integration, $t>0$, are

  \begin{equation}\label{equ:PHt}
    PH(t) = \left\{
           \begin{array}{ll}
             Q_0\cdot e^{\frac{t} {\tau}} \, \big(1 - e^{-\frac{t_{gate}} {\tau}}\big)
             & \hbox{\rm{for}\,\, $t < 0$,} \\
    \vspace{1mm}
             Q_0\cdot \big(1 - e^{-\frac{t_{gate} - t} {\tau}}\big)
             & \hbox{\rm{for}\,\, $ 0 \leq t < t_{gate}, $} \\
             0 & \hbox{\rm{for}\,\, $ t \geq t_{gate}$.}
           \end{array}
         \right.
  \end{equation}

 So far AC coupling has been ignored.
 In the presence of AC\,coupling, the following term has to be subtracted from Eq.\,\ref{equ:PHt}

  \begin{equation}\label{equ:PHACt}
    PH_{AC}(t) = \left\{
           \begin{array}{ll}
    \vspace{1mm}
             Q_0\cdot \Big(
             \frac{e^{\frac{t} {\tau_{AC}}}} {1-\tau / \tau _{AC}} \big(1 - e^{-\frac{t_{gate}} {\tau_{AC}}}\big)-
             \frac{e^{\frac{t} {\tau}}} {(\tau_{AC}/\tau) - 1} \big(1 - e^{-\frac{t_{gate}} {\tau}}\big)
            \Big)
             & \hbox{\rm{for}\,\, $t < 0$,} \\
    \vspace{2mm}
             Q_0\cdot \Big(\frac{1 - e^{-\frac{t_{gate}-t} {\tau_{AC}}}} {1-\tau / \tau _{AC}} -
             \frac{1 - e^{-\frac{t_{gate}-t} {\tau}}} {(\tau_{AC}/\tau) - 1} \Big)
             & \hbox{\rm{for}\,\, $ 0 \leq t < t_{gate}, $} \\
             0 & \hbox{\rm{for}\,\, $ t \geq t_{gate}$.}
           \end{array}
         \right.
  \end{equation}

   \begin{figure}[!ht]
   \centering
   \begin{subfigure}[a]{0.5\textwidth}
   \centering
    \includegraphics[width=0.82\textwidth]{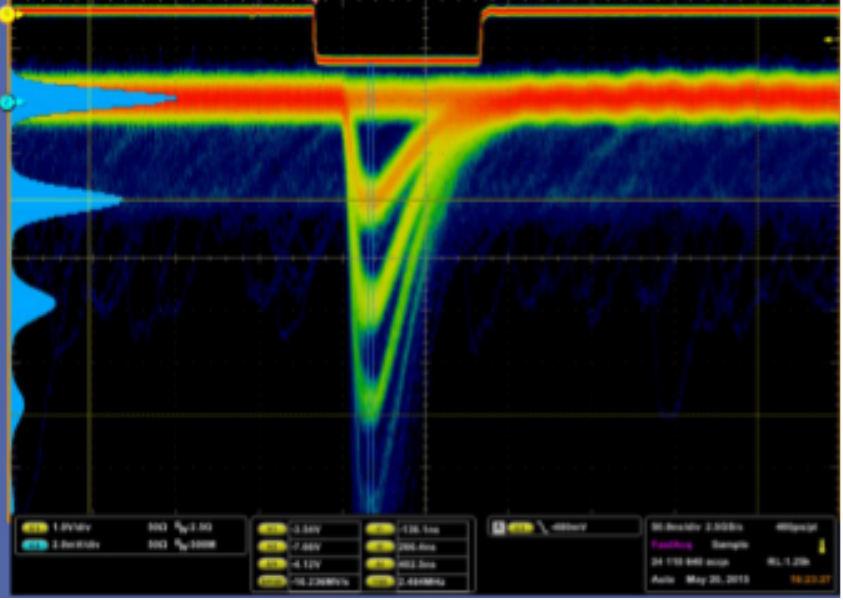}
    \caption{ }
   \end{subfigure}%
    ~
   \begin{subfigure}[a]{0.5\textwidth}
%    \includegraphics[width=\textwidth]{Fig1b.pdf}
%    \includegraphics[width=2.5cm]{Fig2b.pdf}
% DetectorWork\SiPM\Ketek\N-Irradiation\DelayFit02.xmcd
    \includegraphics[width=\textwidth]{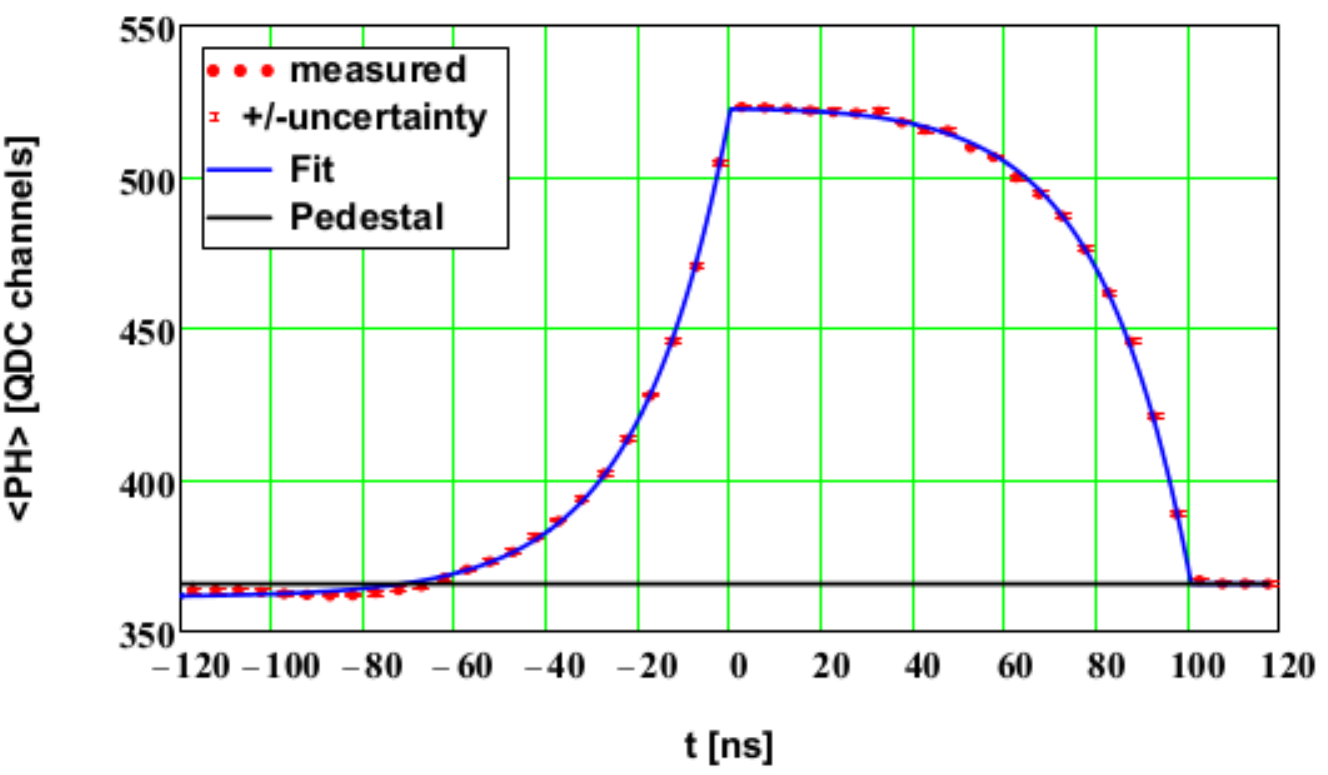}
    \caption{ }
   \end{subfigure}%
   \caption{ (a) Screenshot from the oscilloscope for the SiPM operated at 32\,V illuminated by the pulsed LED.
   Top line, the gate, and bottom lines, the SiPM transients, where bands corresponding to 0, 1, 2, 3, and 4 Geiger discharges can be distinguished.
   (b) Delay curve measured with a pulse-height analyser (CAEN QDC), and the fit described in the text.
%   The value obtained from the fit for the pulse decay time is $13.4 \pm 0.2$\,ns, $35.7 \pm 0.5$\,ns for the start of the gate, $71.1 \pm 0.5$\,ns the gate width and 40\,$\upmu$s for the AC-coupling time constant.
    The line at $PH = 365.5$ indicates the pedestal value.
    If the SiPM pulse precedes the gate (delays < -80\,ns), a smaller $PH$\,value than the pedestal is observed, which is caused by the AC\,coupling of the SiPM readout.}
  \label{fig:Delay}
 \end{figure}

 Fig.\,\ref{fig:Delay}a shows a screenshot of the gate and of SiPM pulses recorded at 32\,V for a SiPM illuminated by the pulsed LED.
 Fig.\,\ref{fig:Delay}b shows the corresponding delay curve, the average $PH$ for the different delay times, $t$, $\langle PH{meas}(t)\rangle $.
 The data points are represented by symbols and the fit by $PH(t)-PH_{AC}(t)$ by the solid line.
 The fit, including the AC coupling, provides a good description of the data, which is not the case if the AC coupling is not taken into account.
% only $PH(t)$ is used.
 The reason is that the base line is higher, if the SiPM signal occurs after the gate ($t > 100$\,ns) than if it occurs before the gate ($t < -80$\,ns), which is caused by the AC\,coupling.
 The horizontal line in Fig.\,\ref{fig:Delay}b shows the base line for a SiPM signal after the gate.

% The results of the fits are shown in Table\,3.
%%% --- Table with tau, tau_AC, gate width, delay and corresponding uncertainties ----

 For a given \textit{PH} threshold, $PH_{thr}$, we define an effective gate width, $t_{gate}^{eff}$, as the time difference between the two intercepts of the delay curve with the constant $PH_{thr}$.
 From Eq.\,\ref{equ:PHt} follows:
 \begin{equation}\label{equ:teff}
  t_{gate}^{eff}(PH_{thr}) = t_{gate} + \tau \cdot \ln \Big(\frac{Q_0-PH_{thr}} {PH_{thr}} \big(1-e^{-\frac{t_{gate}} {\tau }} \big) \Big).
 \end{equation}
 Eq.\,\ref{equ:teff} is used in Sect.\,\ref{sect:PHdark} for determining the dark-count-rate.
 For the parameters determined by the fit to the data, $t_{gate} = (100.67 \pm 0.12)$\,ns and $\tau = (19.95 \pm 0.14)$\,ns,
 $t_{gate}^{eff} (0.5 \cdot Q_0) = (100.54 \pm 0.12)$\,ns, thus essentially equal to $t_{gate}$.
 Only the statistical errors are given, in order to give an idea about the sensitivity of the method.
 No study of systematics has been made, however it is expected that the systematic uncertainties dominate.

  \subsection{Pulse-height measurements with a pulsed LED}
  \label{sect:PHLED}

\begin{figure}[!ht]
   \centering
    \includegraphics[width=0.8\textwidth]{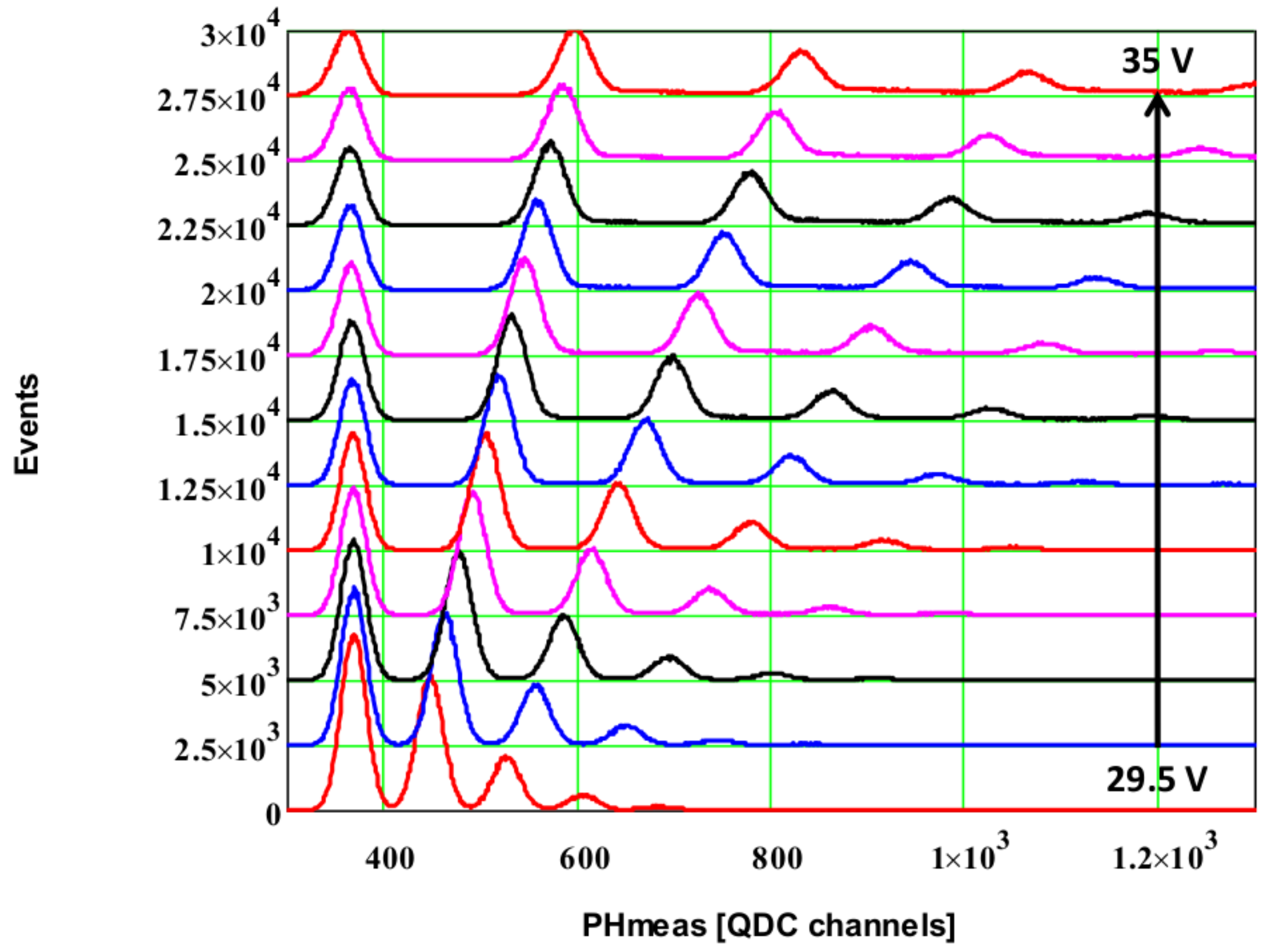}
   \caption{ Pulse-height spectra for voltages from 29.5 to 35\,V in steps of 0.5\,V illuminated by a pulsed LED.
   For clarity the data points are shifted vertically in steps of 2500 for increasing voltages.
   Each data set consists of $5\times 10^5$\,events.}
  \label{fig:Fig6}
% Analysis D:\sync\Junk\DetectorWork\SiPM\Ketek\N-Irradiation\Analysis\MP15\QDC\RadDam15-paper.xmcd
% SiPM: CERN04-W1-B1-S2-MP15V9-dia1mm (N.B. 0 and 10^9 neq)
 \end{figure}

  The \textit{PH} spectra measured with the SiPM illuminated by a pulsed LED, d$N_{LED}/\mathrm{d}PH$, were recorded at 20$^\circ $C and reverse voltages from 29.5 to 35\,V in steps of 0.5\,V, which is above the breakdown voltage of about 27\,V.
  The results are shown in Fig.\,\ref{fig:Fig6}.
  The peaks corresponding to 0, 1, and more Geiger discharges are well separated.
  We follow the usual convention and use the symbol \textit{npe}, the abbreviation of number-of-photoelectrons, for the number of Geiger discharges.
  The linear increase of $Gain$ with voltage is evident from the increase of the distance between the peaks.
  The increase of the photon-detection efficiency, \textit{PDE}, with increasing voltage is seen as a decrease of the fraction of events in the $npe = 0$ peak and an increase of the fraction of events in the $npe \geq 1$ peaks.
  The increase of the widths of the $npe$\,peaks with voltage is due to the increase of the dark-count rate, which via the AC coupling causes additional fluctuation of the baseline.
%  It can be reduced significantly with a readout using correlated double sampling.

 From the $PH$ spectra the voltage dependence of the following SiPM parameters can be obtained:
 the $Gain$ from the distance between the $npe$ peaks,
 the average number of photons initiating a Geiger discharge $\langle N_\gamma \rangle \equiv \mu = -\ln (f_0)$, where $f_0$ is the fraction of events in the pedestal peak corrected for dark pulses,
 the standard deviation of the electronics noise, $\sigma _0$, from the width of the $npe = 0$ peak,
 and the contribution from the gain spread between and in the individual pixels, $\sigma _1$, from the variances of the $npe = i$ peaks using $\sigma _i^2 = \sigma _0^2 + i \cdot \sigma _1^2$.
 In addition, the pulse-height spectra are sensitive to prompt and delayed cross-talk and after-pulsing.
 As the LED light intensity is constant during the voltage scan, $PDE(V) \propto\ \mu (V)$, were \textit{PDE} is the photon-detection efficiency at the wavelength of the LED light.

\begin{figure}[!ht]
   \centering
   \begin{subfigure}[a]{0.5\textwidth}
   \centering
    \includegraphics[width=\textwidth]{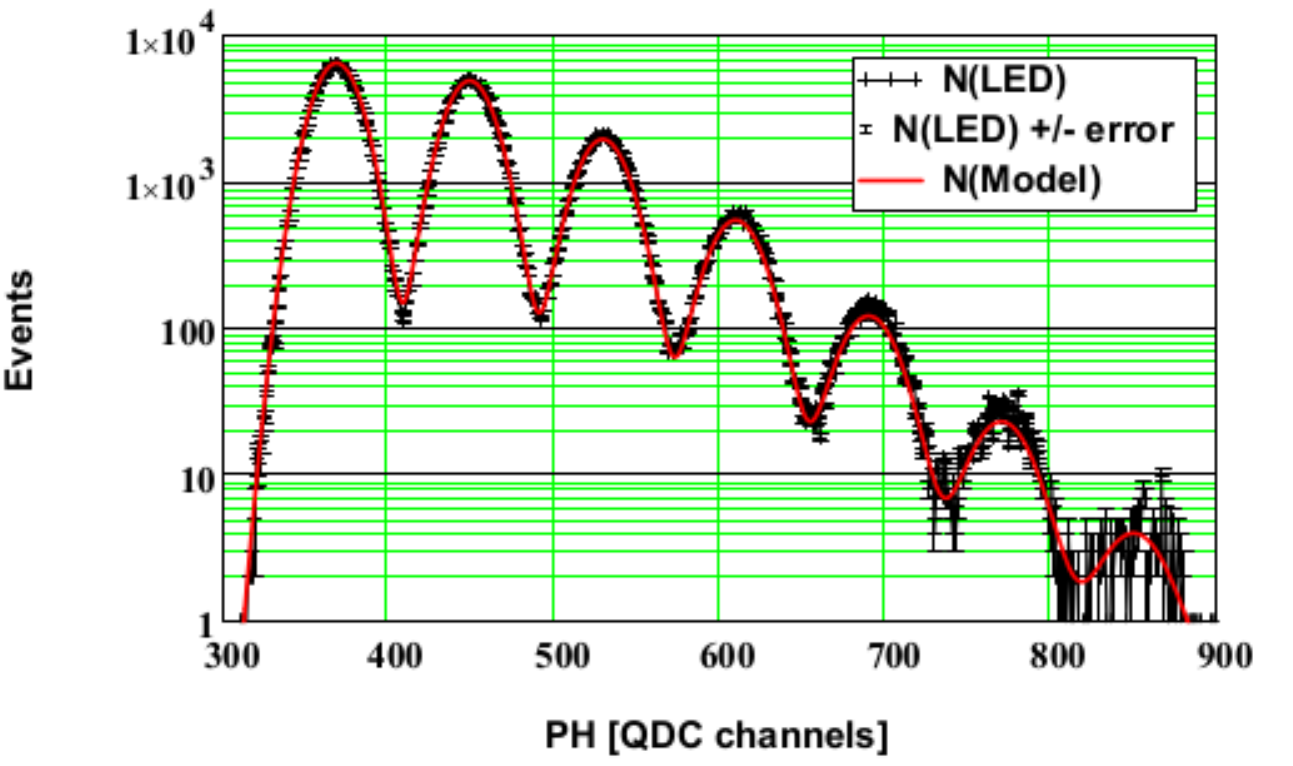}
    \caption{ }
   \end{subfigure}%
    ~
   \begin{subfigure}[a]{0.5\textwidth}
    \includegraphics[width=\textwidth]{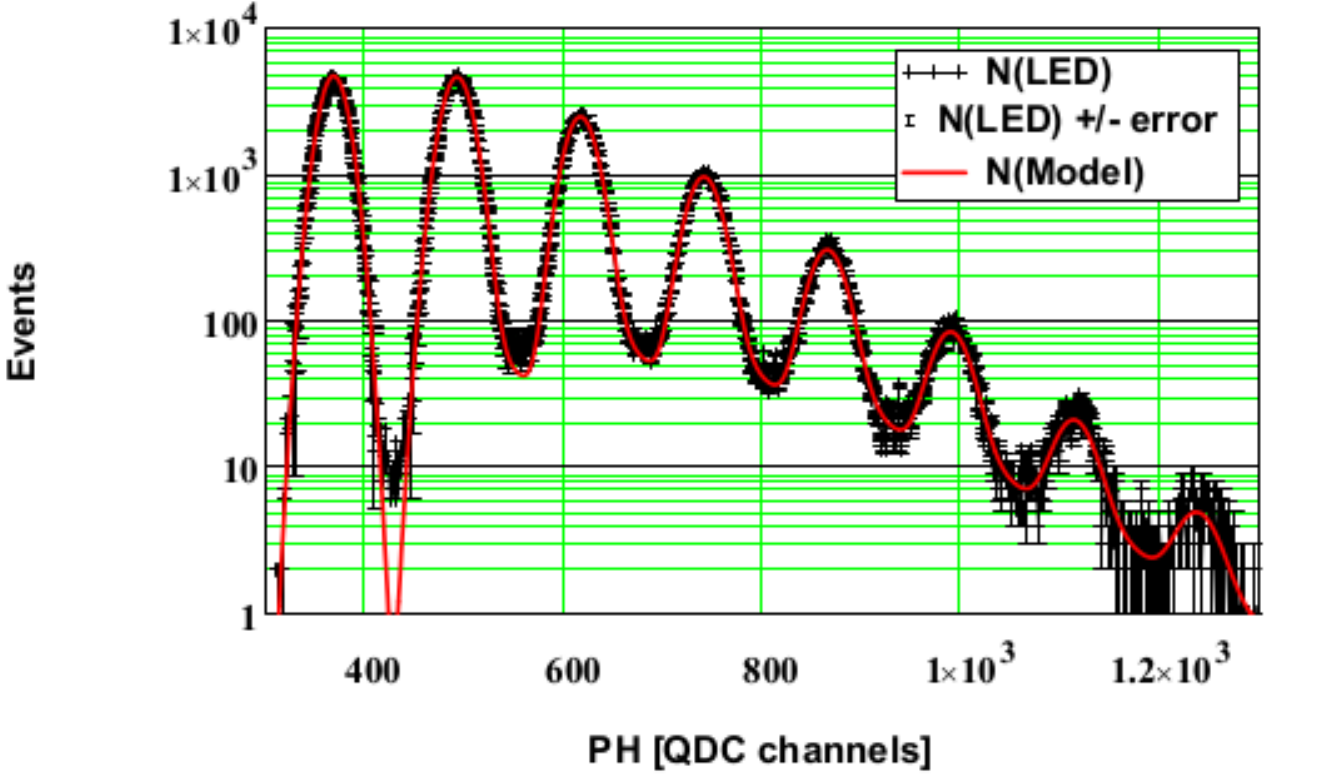}
    \caption{ }
   \end{subfigure}
   ~
   \begin{subfigure}[a]{0.5\textwidth}
   \centering
    \includegraphics[width=\textwidth]{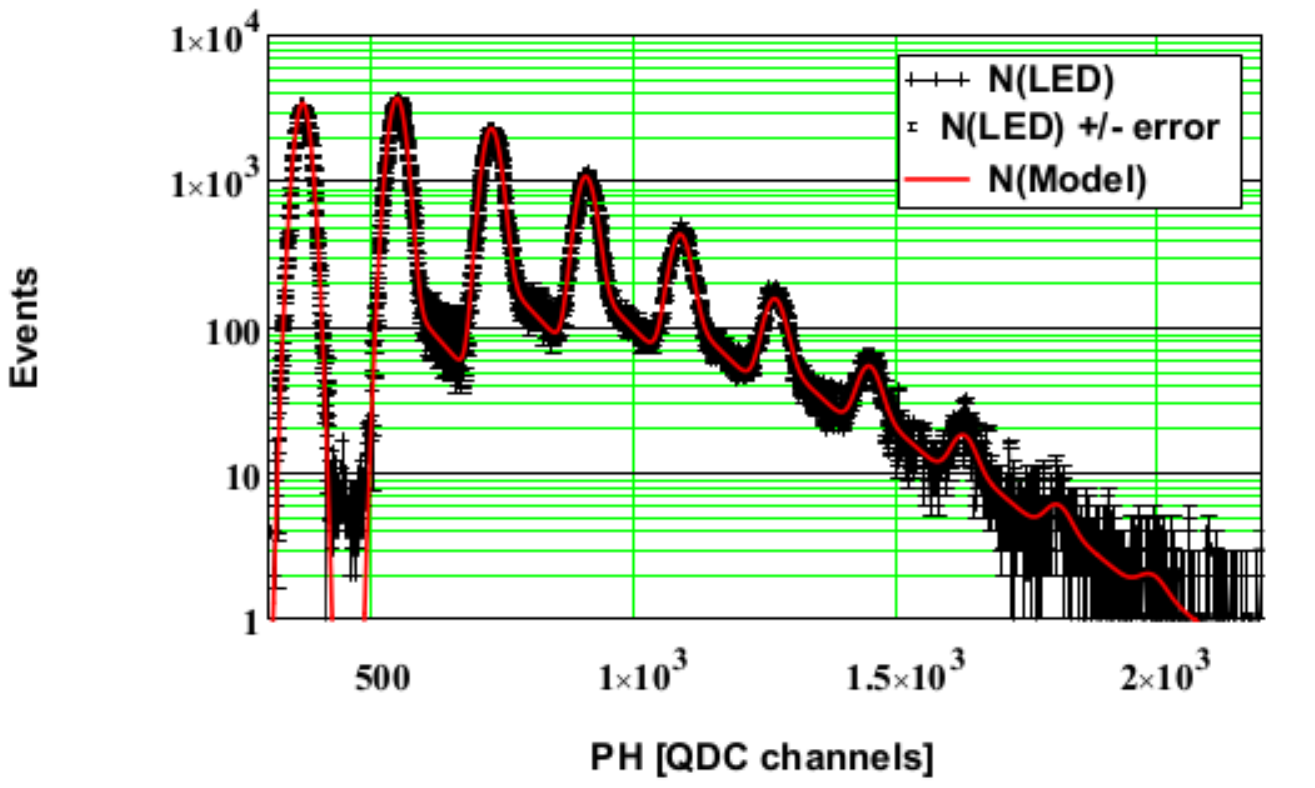}
    \caption{ }
   \end{subfigure}%
    ~
   \begin{subfigure}[a]{0.5\textwidth}
    \includegraphics[width=\textwidth]{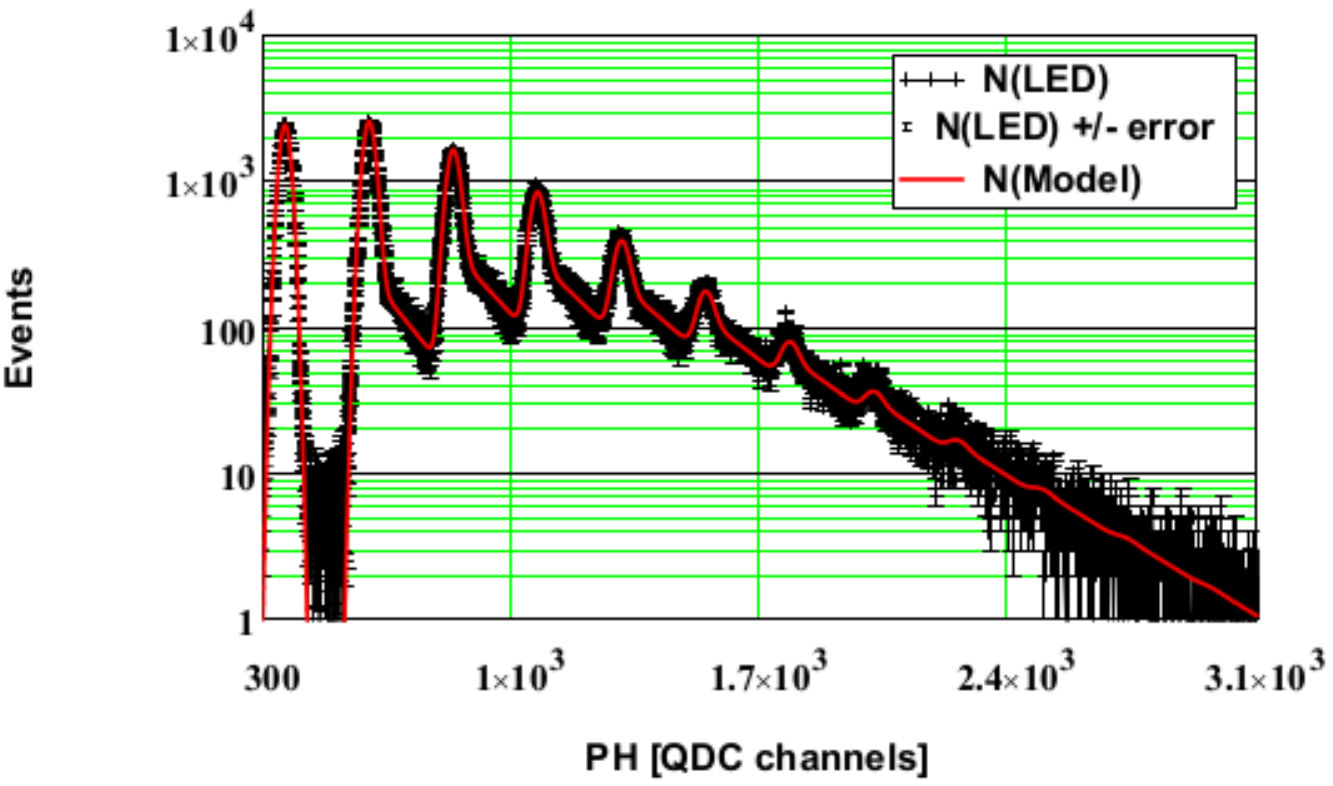}
    \caption{ }
   \end{subfigure}%
   \caption{ Pulse-height spectra in logarithmic scale for the SiPM illuminated by a pulsed LED at (a) 29.5\,V, (b) 31\,V, (c) 33\,V, and 35\,V. The points with error bars are the data, and the solid lines the fits described in the text.
   }
  \label{fig:Fig7}
 \end{figure}

 The standard method of analysing the pulse-height spectra is to fit Gauss functions to the individual $npe$ peaks and derive the quantities discussed above from the positions, widths and number of events in the peaks\,\cite{Eckert:2010,Xu:2014}.
 For the $Gain$ determination, also a Fourier-transform method is frequently used.
 The advantages of these methods is that they are in principle straight-forward.
 However, as seen in Fig.\,\ref{fig:Fig7}, there is a significant number of events in-between the $npe$ peaks, which cannot be described by the Gauss functions.
 Therefore, the regions of the fits have to be selected and it is difficult to estimate how much the parameters extracted are affected.
 In addition, it is not clear if the background in-between the peaks should be subtracted or not.
 We therefore propose a method, which describes the entire \textit{PH} spectrum with a single function.

 We assume that the number of photons initiating a Geiger discharge follows a Poisson distribution with the mean $\mu $.
 Each Geiger discharge can produce secondary Geiger discharges, either prompt or delayed.
 For the prompt cross-talk we assume a Borel distribution with parameter $\lambda $, which results in a Generalized Poisson (\textit{GP}) distribution for $k$\,discharges\,\cite{Consul:1973,Vinogradov:2012}
 \begin{equation}\label{equ:GP}
  GP_{k,\mu, \lambda} = \frac{\mu \cdot (\mu + k \cdot \lambda)^{k-1} \cdot e^{-(\mu + k \cdot \lambda)}} {k !}.
 \end{equation}
 The Borel-branching parameter is denoted $\lambda $ and the probability that a single Geiger discharge produces one or more prompt cross-talk pulses is $1-\exp (- \lambda)$.
 The mean value of the $GP$ distribution is $\mu /(1-\lambda)$ and its variance $\mu / (1-\lambda)^3$.
 We are aware that prompt cross-talk is essentially limited to the neighbours of the discharging pixel, and $GP$ overestimates the additional number of discharging pixels.
 However, as long as $\lambda \lesssim 0.25$, the effect is negligible.
 This has been checked by making the analysis assuming a branching process to the four closest neighbours only.
% We have also used a binomial branching process and found, in accordance with Ref.\,\cite{Vinogradov:2012}, that it results in a worse description of the data.

 Inspecting Fig.\,\ref{fig:Fig7}, which displays the pulse-height spectra measured at 29.5, 31, 33 and 35\,V in logarithmic scale, one notices that with increasing voltage a significant number of events appear in-between the $npe$\,peaks for $npe \geq 1$.
 We attribute them to delayed cross-talk and after-pulsing, named $AP$ in the following.
 From the events  between the $npe = 1$ and $npe = 2$ peaks, which correspond to single-$AP$ events, we conclude that their probability distribution can be approximately described by an exponential
 \begin{equation}\label{EXP1}
   \frac {\mathrm{d}p_{1,1} } {\mathrm{d}PH} = (1 /\beta) \cdot e^{-(PH-PH_1)/\beta}\,\,\mathrm{for}\,PH \geq PH_1,
 \end{equation}
 with the exponential slope $\beta $, and the mean pulse-height of the $npe = 1$ peak $PH_1$.
 Here, and in the following, the first index, $k$, refers to the number of prompt discharges, and the second, $i$, to the number of  $AP$\,discharges.

 A side remark:
 We did not expect an exponential distribution, and actually observe different distributions for other types of SiPMs.
 For a further discussion we refer to Appendix \ref{sect:AppendixA}.
% \ref{sect:Appendix}.

 If there are $i > 1\,AP$\,discharges in an event, which can occur for $k \geq 2$ prompt Geiger discharges, the probability distributions are given by the convolution of $i$ exponentials
 \begin{equation}\label{equ:Conv}
   \frac{\mathrm{d}p_{k,i} }  {\mathrm{d}PH} = \frac{(PH-PH_k)^{i-1}} {(i-1)!\cdot \beta^i  }\, e^{-(PH-PH_k)/\beta} \,\,\mathrm{for}\,PH \geq PH_k,
 \end{equation}
 where $PH_k = Ped + k \cdot Gain$ is the mean \textit{PH} of the $npe=k$ peak.
 The mean pulse-height of the pedestal peak is denoted by $Ped \equiv PH_0$, and the distance between the $npe$ peaks by $Gain$.
 For the probability distribution that $k$ prompt discharges produce $i\,AP$ discharges, the binomial distribution is assumed
 \begin{equation}\label{equ:Binom}
   B_{k,i,\alpha} = \binom{k}{i}\alpha^i(1-\alpha)^{k-i},
 \end{equation}
 with the probability $\alpha $ that a  Geiger discharge also causes an \textit{AP}\,discharge.
 Here a binomial distribution is assumed, as a significant fraction of the \textit{AP}\,discharges is expected  in the same pixel in which the primary discharge has occurred.
 Eq.\,\ref{equ:Binom} ignores the possibility of more than one AP\,discharge in the same pixel.
 This however is a small effect, if the $AP$\,probability does not exceed $\approx 25$\,\%.
 For the prompt Geiger discharges, the noise due to electronics and differences in $Gain$ are taken into account by Gauss functions with variances $\sigma _k^2 = \sigma _0^2 + k \cdot \sigma _1^2$.
 For the smearing of the single-$AP$ exponential from the $npe=k$ peak the approximate formula
 \begin{equation}\label{equ:Exp1}
   \frac {\mathrm{d}p_{k,1}} {\mathrm{d}PH} = \frac{e^{-(PH-PH_k)/\beta} } {\sqrt{2\,\pi}\cdot\sigma_k \cdot \beta } \int\limits_{-\infty}^{PH}\,e^{- \frac{(PH'-PH_k)^2} {2\,\sigma_k^2}} \,\mathrm{d}PH'
 \end{equation}
 is used, which agrees with the exact convolution to within 1\,\% as long as $\sigma_k/\beta < 0.2$.
 The distribution of $i\geq2$ convoluted exponentials is sufficiently smooth, so that no smearing is necessary and Eq.\,\ref{equ:Conv} can be used.
% No smearing is necessary for the  $i\geq2$ convoluted exponentials, and Eq.\,\ref{equ:Conv} is used in this case.
 Thus the complete probability density distribution reads:
  \begin{equation}\label{equ:Model}
   \frac {\mathrm{d}p } {\mathrm{d}PH} = \left\{
       \begin{array}{ll}
        \Big[GP_{0, \mu ,\lambda} \cdot Gauss(PH; 0, \sigma _0)\,  + \\
        \sum_{k=1}^{k_{max}} \Big(GP_{k,\mu,\lambda} \cdot
        \big(B_{k,0,\alpha} \cdot Gauss(PH;k,\sigma_k)\, + \\
        B_{k,1,\alpha} \cdot \frac{\mathrm{d}p_{k,1}(PH)} {\mathrm{d}PH} +
        \sum_{i=2}^{k} B_{k,i,\alpha} \cdot \frac{\mathrm{d}p_{k,i}(PH)} {\mathrm{d}PH}
        \big) \Big) \, \Big],
           \end{array}
         \right.
  \end{equation}
 with
 \begin{equation}\label{Gauss}
   Gauss(PH;k,\sigma_k) =  \frac{1}{\sqrt{2\,\pi}\,\sigma_k}  \cdot e^{-\frac{PH-(Ped + k\cdot Gain)} {2\,\sigma_k^2}}.
 \end{equation}
 This probability density distribution, multiplied with the normalisation constant, \textit{Norm}, is fitted to the $PH$ spectra to determine the best estimate of the 9 parameters:
 the normalisation \textit{Norm},
 the mean number of photons initiating a Geiger discharge $\mu$,
 the branching probability for prompt cross talk $\lambda$,
 the after-pulsing probability $\alpha$,
 the inverse of the exponential slope of the after-pulse \textit{PH} distribution $\beta$,
 as well as \textit{Ped}, \textit{Gain}, $\sigma_0$, and $\sigma_1$.
 The value of $k_{max}$ for the fit of the low-light data is set to 8.

 The results of the fits, evaluated for $k_{max} = 15$, are shown as solid lines in Fig.\,\ref{fig:Fig7} for the data at 29.5, 31, 33, and 35\,V.
 The description of the data is satisfactory, with $\chi ^2/NDF$ values between 1.0 and 1.8 for numbers of degrees of freedom, \textit{NDF}, between 460 and 1600.
% For the data shown in Fig.\,\ref{fig:Fig7}, the light intensity of the LED has been adjusted to result in $\mu \approx 1$ for 30\,V, which allows a precise determination of the SiPM parameters.

      \begin{figure}[!ht]
%Analysis: D:\sync\Junk\DetectorWork\SiPM\Ketek\NIrradiation\Analysis\MP15\QDC\QDC-Results03.xmcd
%
   \centering
   \begin{subfigure}[a]{0.5\textwidth}
   \centering
    \includegraphics[width=\textwidth]{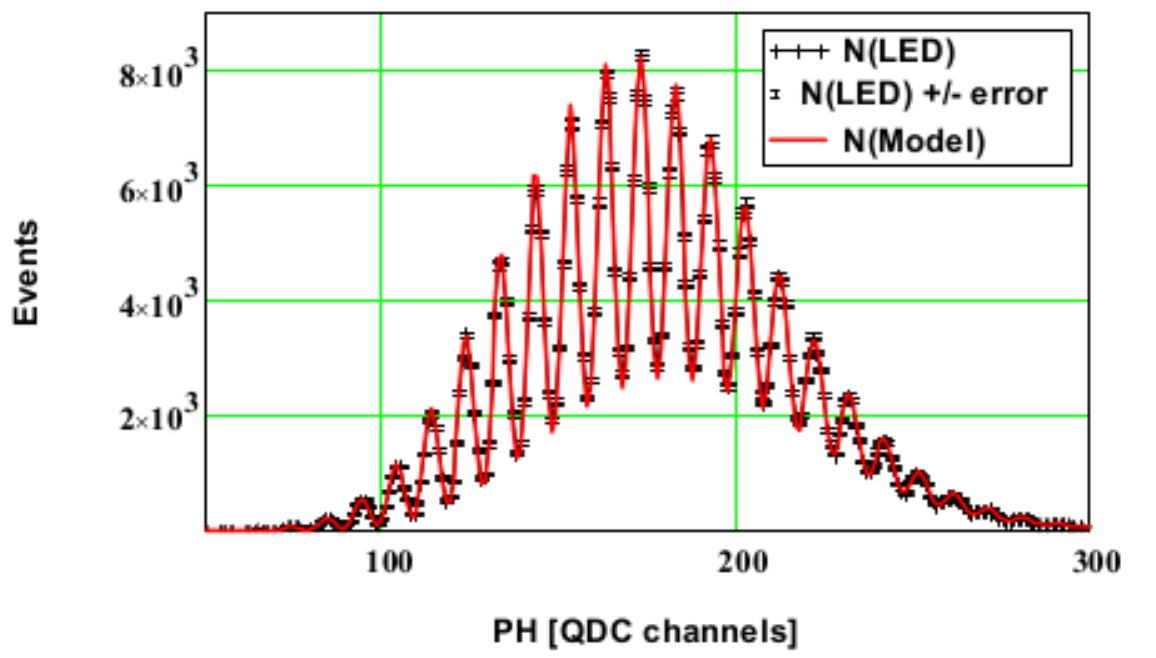}
    \caption{ }
   \end{subfigure}%
    ~
   \begin{subfigure}[a]{0.5\textwidth}
    \includegraphics[width=\textwidth]{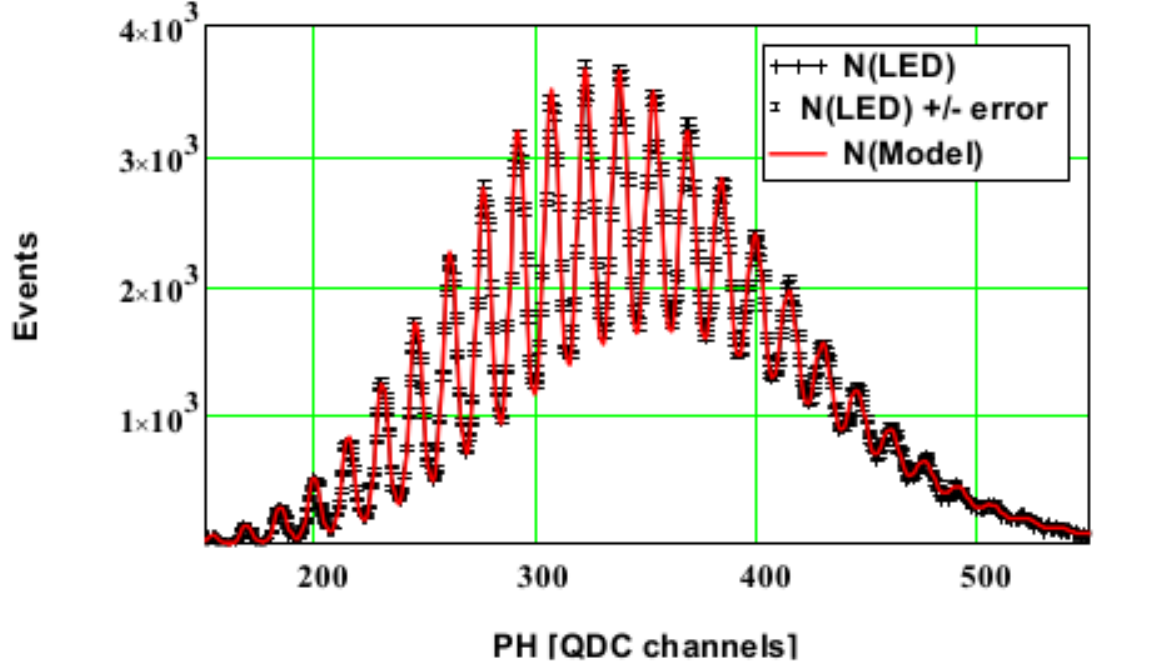}
    \caption{ }
   \end{subfigure}
   ~
   \begin{subfigure}[a]{0.5\textwidth}
   \centering
    \includegraphics[width=\textwidth]{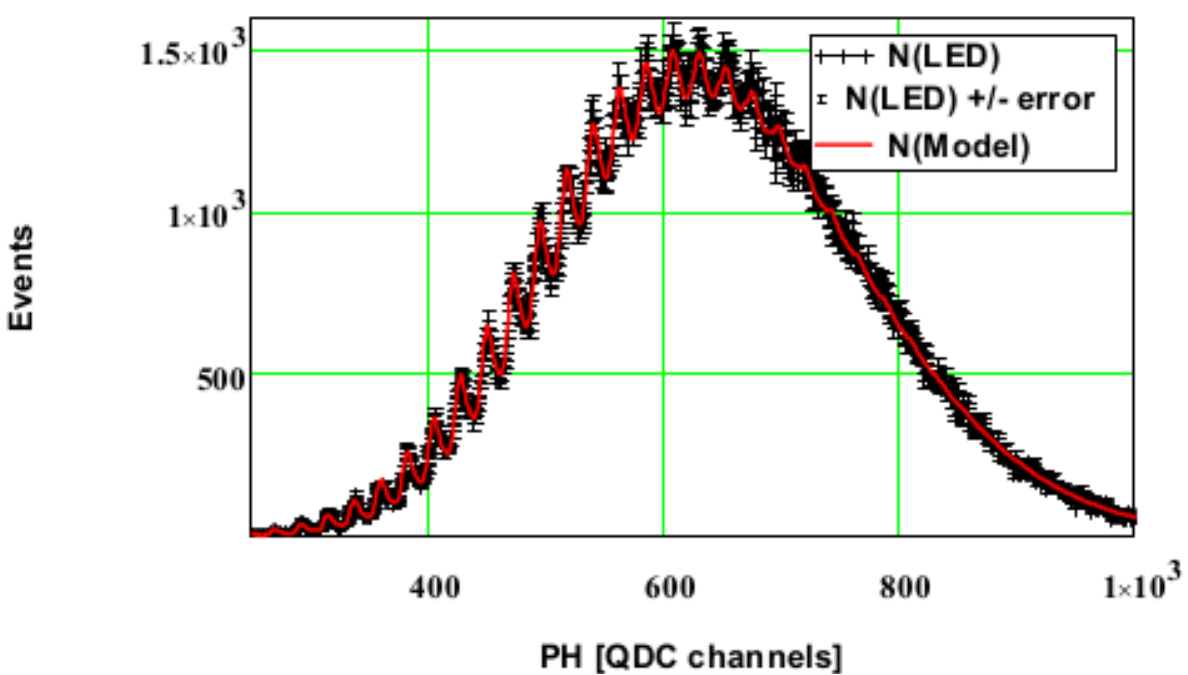}
    \caption{ }
   \end{subfigure}%
    ~
   \begin{subfigure}[a]{0.5\textwidth}
    \includegraphics[width=\textwidth]{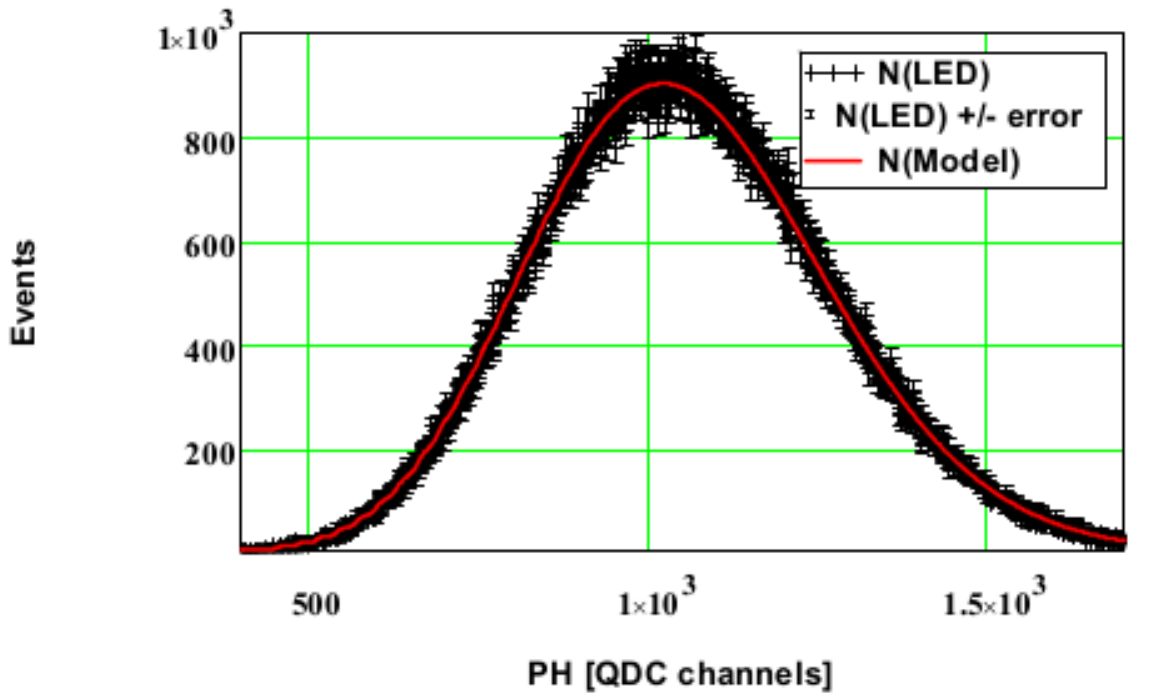}
    \caption{ }
   \end{subfigure}%
   \caption{ Pulse-height spectra for the SiPM illuminated by the pulsed LED for a 16 times higher light intensity compared to the data shown in Fig.\,\ref{fig:Fig7} at (a) 29.5\,V, (b) 31\,V, (c) 33\,V, and 35\,V. The points with error bars are the data, and the solid lines the results of the model calculations using Eq.\,\ref{equ:Model} with the SiPM parameters determined from the fits for low light intensities and $k_{max} = 50 $.
   }
  \label{fig:Fig8}
 \end{figure}

 \begin{figure}[!ht]
%Analysis: D:\sync\Junk\DetectorWork\SiPM\Ketek\NIrradiation\Analysis\MP15\QDC\QDC-Results03.xmcd
%
   \centering
    \includegraphics[width=\textwidth]{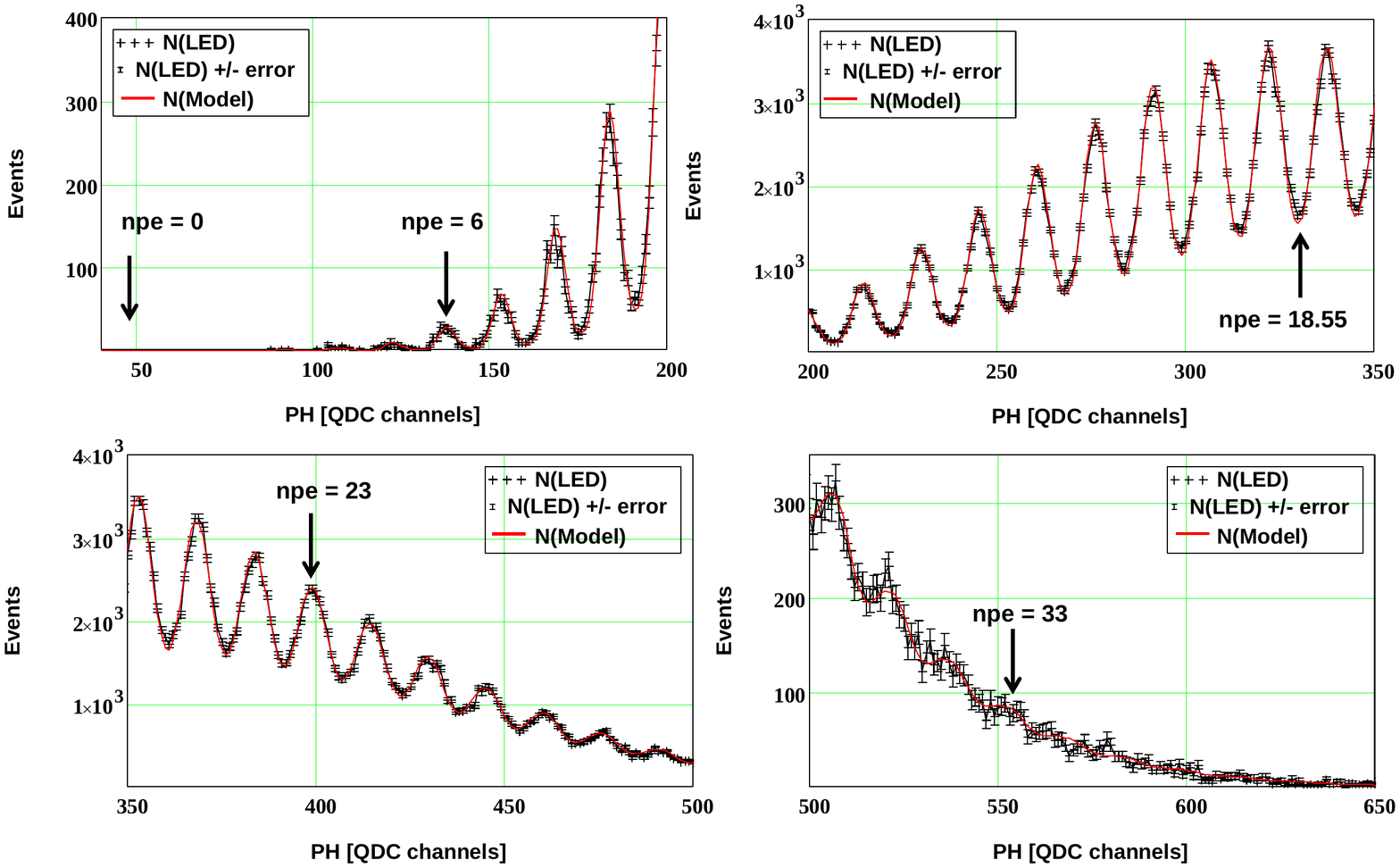}
   \caption{ Expanded view of the pulse-height spectrum for the SiPM at 31\,V already shown in Fig.\,\ref{fig:Fig8}b, which demonstrates the quality of the prediction using the results from the fit to the data with low-light intensity. Arrows indicate the pedestal, $npe =0$, the average number of photoelectrons, $npe = \mu = 18.55$, and the peaks corresponding to $npe = 6$, 23 and 33.
   }
  \label{fig:Fig9}
 \end{figure}

 In addition, data with an approximately 16 times higher LED intensity have been taken.
 To avoid saturation  the low-gain QDC channel, with an $\approx 1/8$ lower gain, has been used.
 One aim of these measurements was to verify, if the parameters determined from the low-light data can be used to predict the SiPM response for higher light intensities.
 Fig.\,\ref{fig:Fig8} shows the comparison of the prediction with the data for the voltages of 29.5, 31, 33, and 35\,V.
 The agreement is excellent.
 This can be better judged from Fig.\,\ref{fig:Fig9}, which shows as an example the 31\,V data with an expanded scale.
 Arrows indicate the positions of $npe = 0$ (pedestal), 6, 23 and 33.
% ,  and of $\mu = 18.55$ determined from the fit.
 For the prediction, the values of \textit{Norm} is adjusted for every voltage.
% The values are close to, but not identical with the number of entries.
 The values differ by up to 1\,\% from the number of entries.
 The reason is that in the comparison of the data with the model prediction, the value of the function and not the integral over the bin is used, which causes small differences.
 The values of $\lambda $ and $\alpha $ are taken from the low-light fit.
 To account for the increased light intensity, the low-light value of $\mu $  is multiplied by the voltage-independent factor 16.275, which results in the best description of the data.
 The value for the 31\,V data, $\mu = 18.55$, is marked by an arrow in Fig.\,\ref{fig:Fig9}.
 To account for the reduced QDC gain, the low-light \textit{Gain} value is multiplied by 1/7.85, again determined from the model-to-data comparison.
 For the evaluation of Eq.\,\ref{equ:Model} $k_{max} = 50$ is used.

      \begin{figure}[!ht]
% DetectorWork/SiPM/Ketek/N-Irradiation/Analysis/MP15/QDC/QDC-Results02-paper.pdf
   \centering
   \begin{subfigure}[a]{0.5\textwidth}
   \centering
    \includegraphics[width=\textwidth]{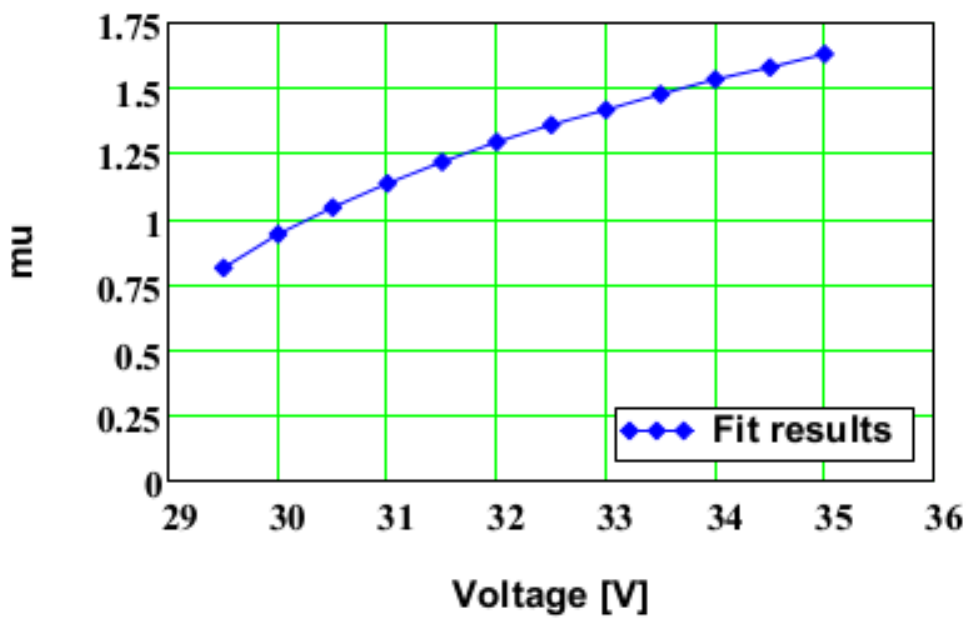}
    \caption{ }
   \end{subfigure}%
    ~
   \begin{subfigure}[a]{0.5\textwidth}
    \includegraphics[width=\textwidth]{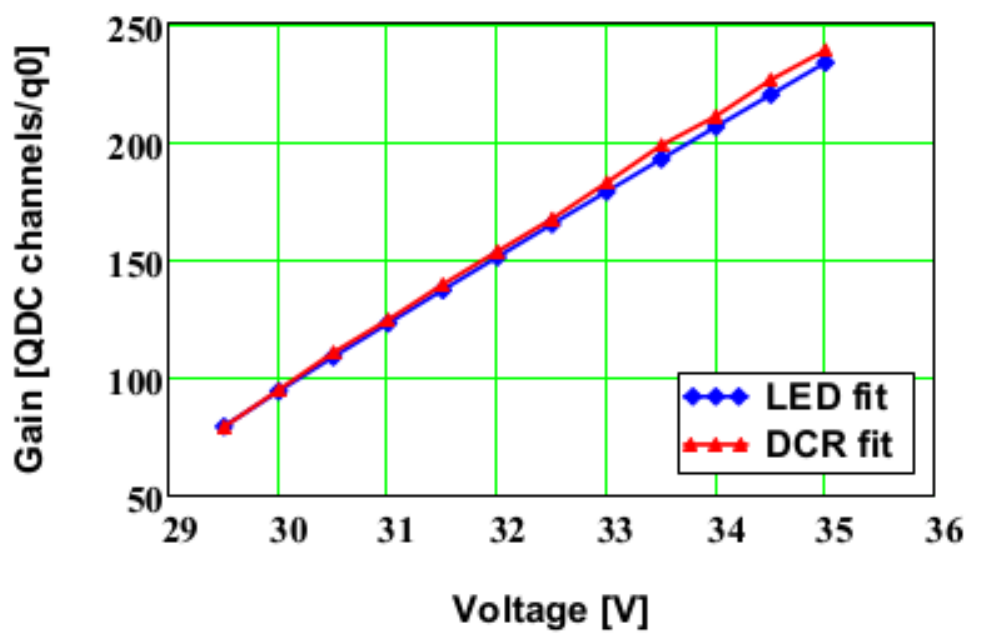}
    \caption{ }
   \end{subfigure}
   ~
   \begin{subfigure}[a]{0.5\textwidth}
   \centering
    \includegraphics[width=\textwidth]{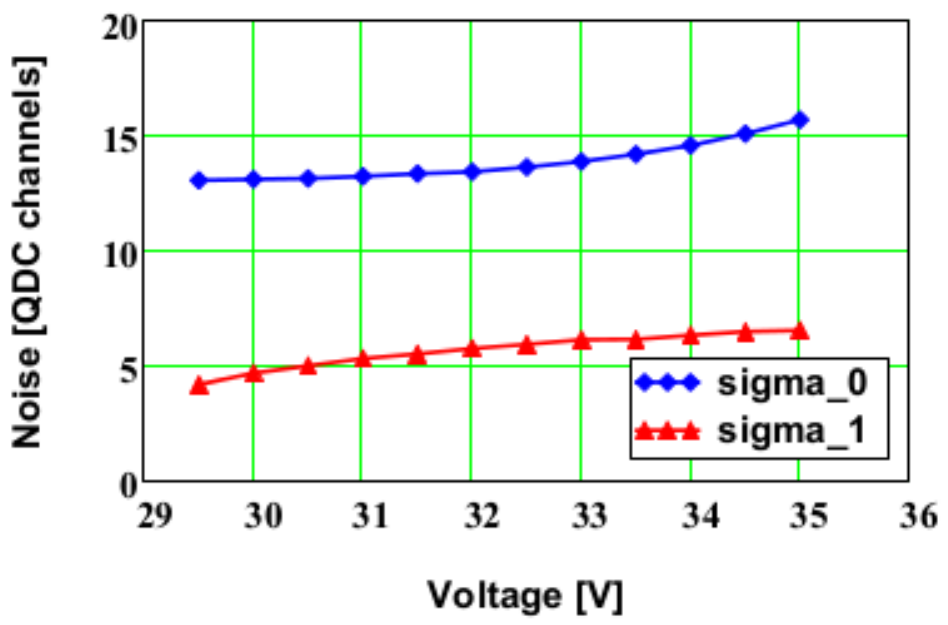}
    \caption{ }
   \end{subfigure}%
    ~
   \begin{subfigure}[a]{0.5\textwidth}
    \includegraphics[width=\textwidth]{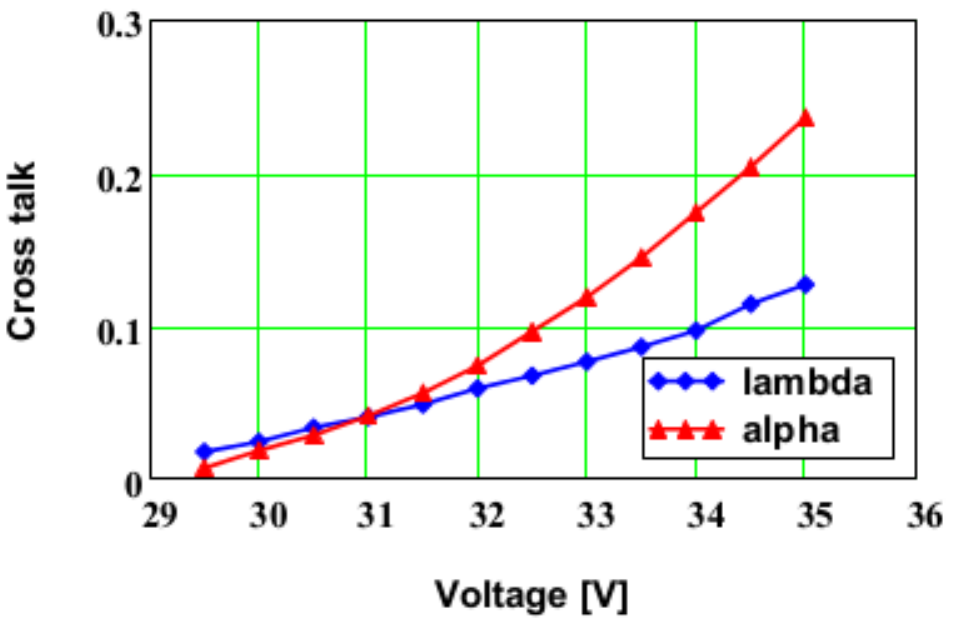}
    \caption{ }
   \end{subfigure}%
   \caption{ Voltage dependence of (a) $ \mu $ from the low-light fits,
    (b) \textit{Gain} in units of QDC channels over elementary charge $q_0$ from the low-light fits and the \textit{DCR} fits presented in Sect.\,\ref{sect:PHdark}, and
    (c) $\sigma_0$, $\sigma_1$ and
    (d) $\alpha$ and $\lambda $   from the low-light fits.
   }
  \label{fig:FigResults}
 \end{figure}

 Fig.\,\ref{fig:FigResults} shows the voltage dependence of $\mu $, \textit{Gain}, $\sigma_0$, $\sigma_1$, $\alpha$ and $\lambda $ for the fits to the low-light data.
 The statistical errors from the fits are smaller than the size of the dots.
 The increase with voltage of $\mu $ reflects the increase of the photon-detection efficiency at the wavelength of the LED light of 470\,nm.
 As expected, the gain increases linearly with voltage with a slope of $\mathrm{d}Gain/\mathrm{d}V = 28.01 \pm 0.05$\, QDC\,channnels/V, and an intercept of $V_{to} = 26.638 \pm 0.020 $\,V, the voltage, at which the Geiger discharges turns off.
 We note that for the SiPM studied, the breakdown voltage determined from the current-voltage characteristics is about 1\,V above $V_{to}$\,\cite{Chmill:2016}.
 The width of the pedestal peak, $\sigma_0 $ is constant up to 32.5\,V, and then shows a small increase, which is caused by the increase in dark-count rate and \textit{Gain}, and the AC coupling of the readout.
 %The value of $\sigma_1$, which in the model is related to the pixel-to-pixel and in-pixels gain variations, also shows a small increase, for which we do not have an explanation.
  The value of $\sigma_1$, which describes the variations of \textit{Gain} between pixels and within pixels, also shows a small increase.
  As $Gain = \mathrm{d}Gain/\mathrm{d}V \cdot ( V - V_{to} ) \propto C_{pix} \cdot ( V - V_{to} )$, $\sigma_1$ has
  one contribution $\sigma_1^{(1)} = ( V - V_{to} )\cdot \delta C_{pix}$ from the variation of the pixel capacitance, $\delta C_{pix}$, and
  a second contribution $\sigma_1^{(2)} = \mathrm{d}Gain/\mathrm{d}V \cdot \delta V_{to}$ from the variation of the threshold voltage, $\delta V_{to}$.
  The term $\delta C_{pix}$ is expected not to depend on \textit{V}, and from the weak voltage dependence of $\sigma _1$ we can exclude a significant contribution from the first term, which is $\propto (V - V_{to})$.
  We conclude that the second term, $\sigma_1^{(2)}$, dominates, and, using the value of $\mathrm{d}Gain/\mathrm{d}V$ from Fig.\,\ref{fig:FigResults}b, we find that $\delta V_{to}$ changes from 150\,mV at 29.5\,V to 230\,mV at 35\,V.
%  From the slow variation of $\sigma_1 $ shown in Fig.\,\ref{fig:FigResults}c, we conclude that the second contribution dominates, and using the value of $\mathrm{d}Gain/\mathrm{d}V$ from Fig.\,\ref{fig:FigResults}b we find that $\delta V_{to}$ changes from 150\,mV at 29.5\,V to 230\,mV at 35\,V.
%  The first contribution is expected to result from variations of the pixel capacitances, $\delta C_{pix}$, resulting in $\sigma_1 =  (V - V_{to}) \cdot \delta C_{pix}$, which increases with voltage.
%  The voltage at which the Geiger discharge turns off, is denoted $V_{to}$.
%  The second contribution, due to fluctuations of $V_{to}$, $\delta V_{to}$, gives $\sigma_1 = C_{pix} \cdot \delta V_{to}$.
%  To determine the voltage dependence of $\delta V_{to}$ requires a detailed simulation of the Geiger discharge, which is beyond the scope of this paper.
%  However, we expect only a weak voltage dependence, and thus an approximately constant $\sigma _1$.
%  From the moderate voltage dependence of $\sigma _1$, we conclude that the fluctuations due to $\delta V_{to}$ dominate.
% It may reflect a non-perfect modeling of the pulse-height spectra.
 Finally, Fig.\,\ref{fig:FigResults}\,d shows the voltage dependence of the prompt-cross-talk probability, $\lambda $ and the after-pulse probability $\alpha $.
 As expected, they both increase with voltage.
% For a comparison with the determination of correlated noise from the measurements without illumination, we refer to Sect.\,\ref{sect:PHdark}.

 \emph{To summarise:}
 The model developed provides with a single function a precise description of the entire SiPM pulse-height spectra for pulsed light.
 Both low-light and high-light data are described with the same SiPM parameters, which is a significant test of the model.
 From the correct description of the intensities of the \textit{npe} peaks, we conclude that the Borel-branching process for the prompt cross-talk, which results in a Generalised Poisson distribution of the number of prompt Geiger discharges, is a valid assumption.
 This conclusion has already been reached in Ref.\,\cite{Vinogradov:2012}.
 From the correct description of the smooth background, we conclude that the assumed exponential pulse-height distribution and the binomial-branching process for the after-pulses are also valid.
 We expect that the pulse-height distribution for after-pulses will be different for  different SiPM designs, and thus has to be determined for each case.
 An example of a possible parametrisation is derived in Appendix\,A.

 Compared to previous methods, the entire pulse-height spectra are fitted by a single function and no bin selection is required.
 This significantly eases a reliable and fully automatic SiPM characterisation.
% The validity of the model is also demonstrated by correctly predicting the pulse-height spectra for high-light intensities from the parameters obtained from the fits to the low-light spectra.
 In Sect.\,\ref{sect:Application} the complete description of the pulse-height spectra will be used to determine the optimal SiPM-operating conditions and develop and test a new method to determine $\mu $ and \textit{Gain}, when the \textit{npe} peaks can not be separated for the \textit{Gain} determination and the methods available so far fail.

  \subsection{Pulse-height measurement without illumination}
  \label{sect:PHdark}

\begin{figure}[!ht]
%D:\sync\Junk\DetectorWork\SiPM\Ketek\NIrradiation\Analysis\MP15\QDC\QDC-150827\Dark-Fit-03.xmcd
   \centering
   \begin{subfigure}[a]{0.5\textwidth}
   \centering
    \includegraphics[width=\textwidth]{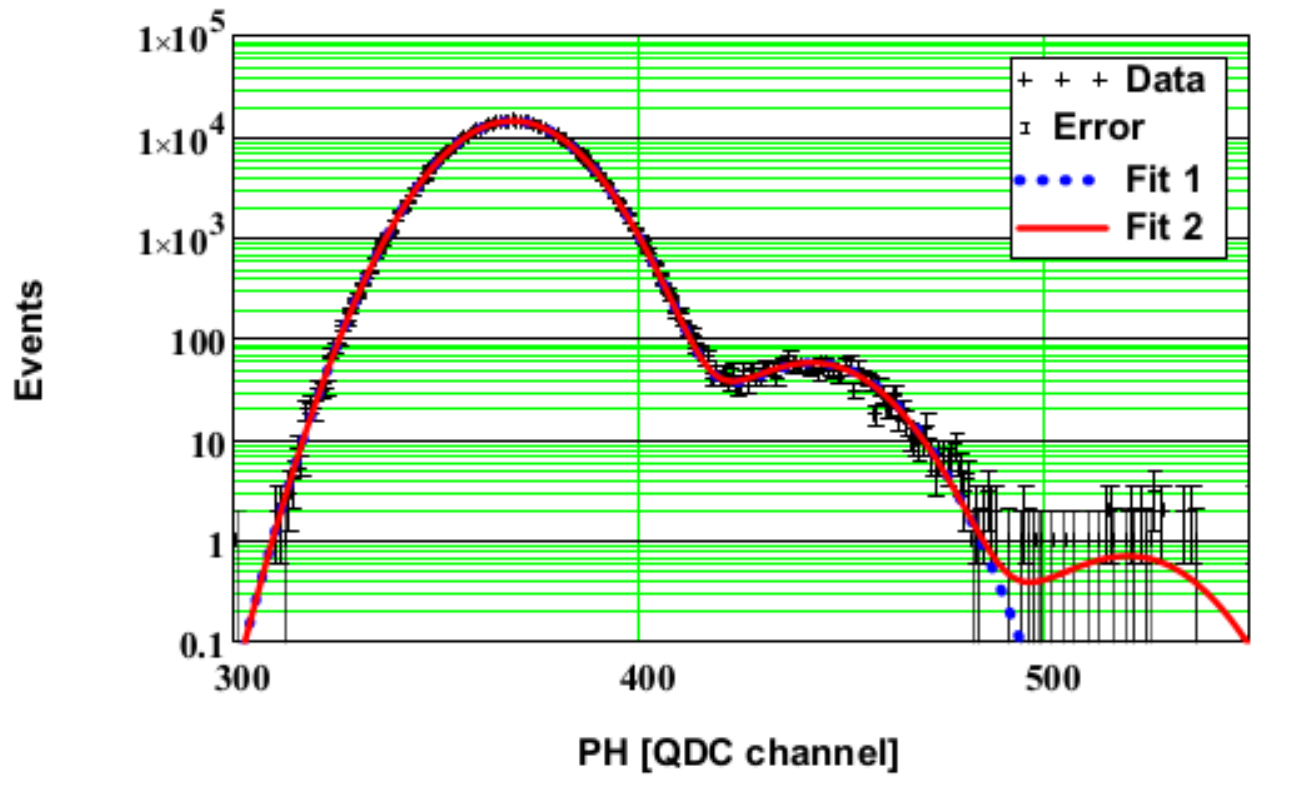}
    \caption{ }
   \end{subfigure}%
    ~
   \begin{subfigure}[a]{0.5\textwidth}
    \includegraphics[width=\textwidth]{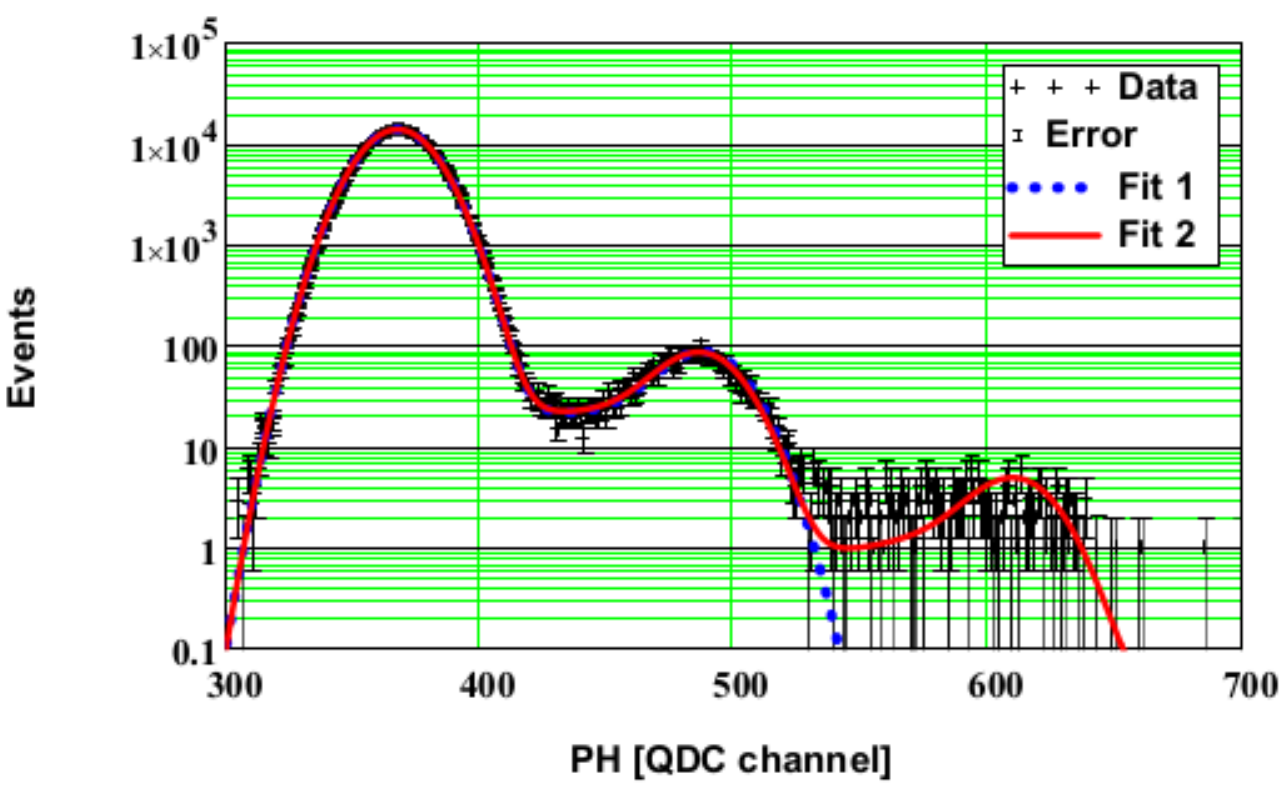}
    \caption{ }
   \end{subfigure}
   ~
   \begin{subfigure}[a]{0.5\textwidth}
   \centering
    \includegraphics[width=\textwidth]{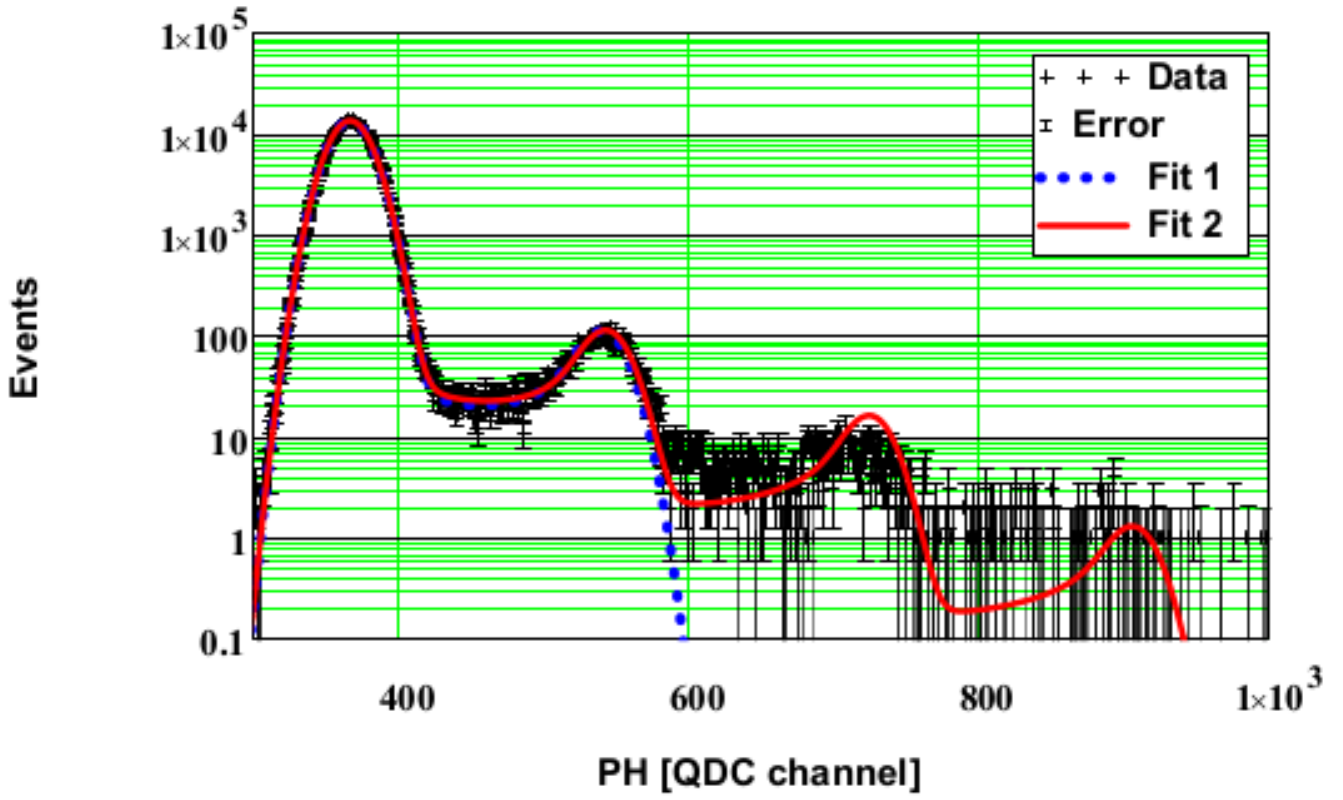}
    \caption{ }
   \end{subfigure}%
    ~
   \begin{subfigure}[a]{0.5\textwidth}
    \includegraphics[width=\textwidth]{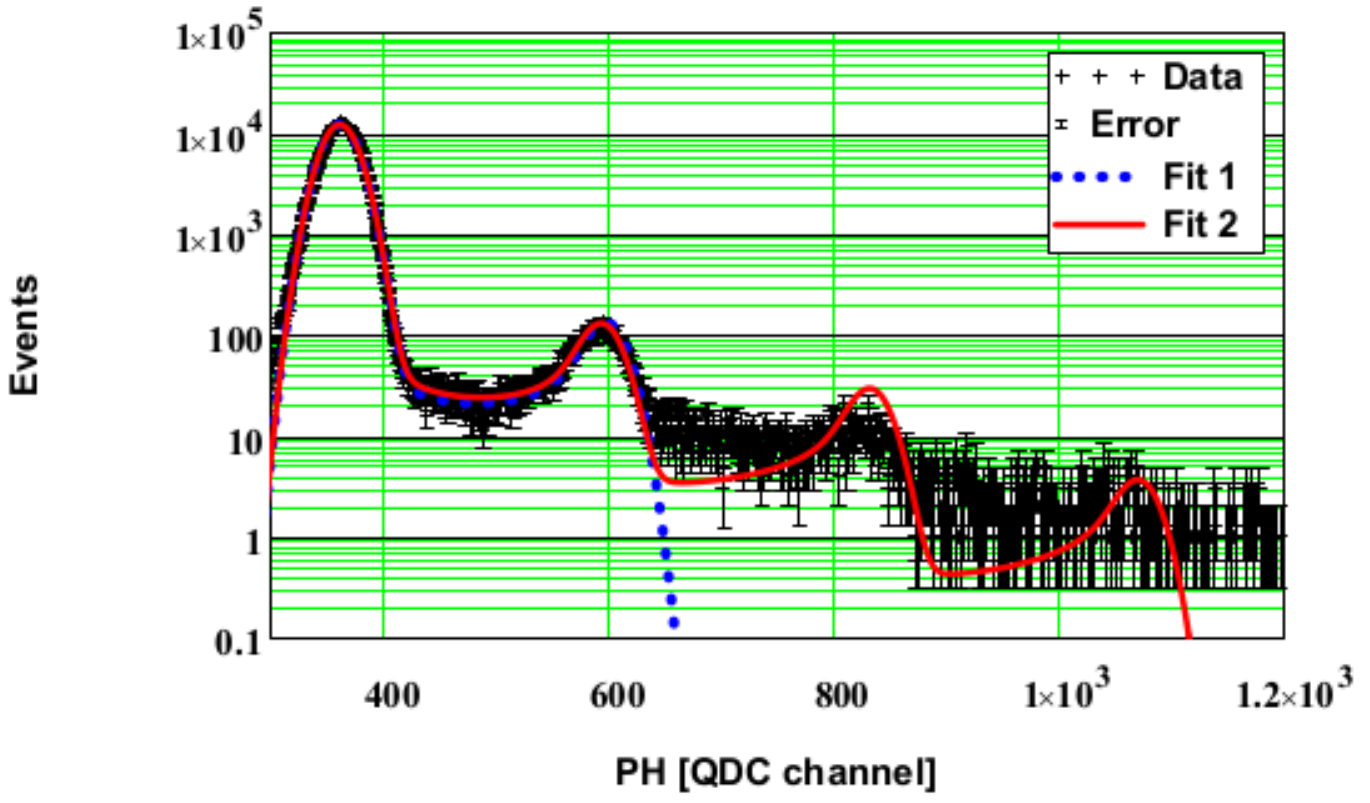}
    \caption{ }
   \end{subfigure}%
   \caption{ Pulse-height spectra measured without illumination with the results of the fits discussed in the text for
    (a) 29.5\,V,
    (b) 31.0\,V,
    (c) 33.0\,V, and
    (d) 35.0\,V.
   The dotted line is the model with single dark pulses only, the solid line the model, which includes multiple dark pulses and prompt cross-talk.}
  \label{fig:FigDCR}
 \end{figure}

  Pulse-height spectra measured without illumination, d$N_{dark}/\mathrm{d}PH$, were recorded at $ 20^{\,\circ }$C for reverse voltages from 29.5 to 35\,V in steps of 0.5\,V.
  The spectra were taken using the gate from a pulse generator, so that gate and dark pulse are uncorrelated in time.
%   which is above the breakdown voltage of about 27\,V.
  As examples, Fig.\,\ref{fig:FigDCR}a shows the \textit{PH} spectra at 29.5, 31, 33, and 35\,V.
 The \textit{PH} spectra are dominated by the pedestal peak, corresponding to no Geiger discharge, $npe = 0$, a smaller peak corresponding to one Geiger discharge, $npe = 1$, caused by a single dark pulse overlapping with the gate, and at higher operating voltages, entries beyond the $npe = 1$ peak, caused by multiple dark counts and correlated noise pulses.
 The entries in-between the $npe = 0$ and the $npe =1$ peak are due to Geiger-discharge pulses, which only partially overlap with the QDC gate.
 For measurements in which gate and dark pulse are correlated in time, which can be realised if transients are recorded, dark pulses with partial overlap with the gate can be avoided.
 This significantly simplifies the analysis and is very suitable for an automated analysis as presented in Ref.\,\cite{Piemonte:2012}.

 The standard way to analyse the $PH_{dark}$ spectra\,\cite{Eckert:2010, Xu:2014} is to first determine $PH_0$, the position of the $npe = 0$ peak using a fit by a Gauss function and take \textit{Gain} from the analysis of the low-light spectra.
%The standard way to analyse the $PH_{dark}$ spectra\,\cite{Eckert:2010, Xu:2014} is to first determine $PH_0$ and $PH_1$, the positions of the $npe = 0$ and the $npe = 1$ peaks using fits by Gauss functions, and calculate the gain, $Gain = PH_1 - PH_0$.
% The $PH_{dark}$\,spectra, as well as the spectra with illumination, $PH_{LED}$, can be used for the gain determination (see Sect.\,\ref{sect:PHLED}).
% The latter is preferred, as it has the higher accuracy.
% The gain in channel numbers is $Gain = PH_1 -PH_0$.
 Then, the fraction of events with $PH > (PH_0 + 0.5\,Gain)$, called $f_{0.5}$, and the fraction above $(PH_0 + 1.5\,Gain),\, f_{1.5}$, are determined.
 The dark-count rate is calculated using $DCR = f_{0.5} / t_{gate}$, and the correlated noise using $CN = f_{1.5}/f_{0.5}$.
 In order to obtain reliable results, the $npe = 0$ and the $npe = 1$\,peaks have to be sufficiently separated, which, e.g. is not the case for the pulse-height spectrum at 29.5\,V.
 However, this can be corrected by subtracting from $f_{0.5}$ the tail of the $npe = 0$ Gauss function.
 Just subtracting the number of events in the $npe = 0$ peak from the total number of events does not take into account the fraction of events with $PH$s between the pedestal and 1/2 the $npe = 0$ peak and thus overestimates the dark-count rate.
 In this case the effective gate width, defined in Eq.\,\ref{equ:teff}, for the corresponding pulse-height threshold should be used.

 For deriving the shape of the \textit{PH} spectrum for a single dark-count pulse, we refer to Fig.\,\ref{fig:Delay}b, which shows the mean \textit{PH} integrated during the gate as a function of the time difference between the Geiger discharge and the start of the integration by the gate.
 Dark-count pulses are distributed randomly in time and their rate is \textit{DCR}.
 The probability of pulses in the interval between \textit{PH} and \textit{PH}+d\textit{PH} is proportional to the time interval $\mathrm{d}t = \mathrm{d}PH/|\mathrm{d}PH/\mathrm{d}t|$.
 For a normalised current pulse: $I_{norm} (t') = 1/\tau \cdot e^{-t'/\tau}$ for $t' \geq 0$ and $I(t') = 0 $ for $t' < 0$,  $PH_{norm}$ for a dark pulse at the time $t$ relative to the start of the integration by the gate is given by Eq.\,\ref{equ:PHt} with $Q_0 = 1$:

% The method proposed here, tries to model the entire $PH_{dark}$ spectrum.
% It takes into account that the dark counts occur randomly in time and that only a fraction of them overlaps fully with the gate, which results in entries outside of the $npe$\,\,peaks.
% To explain the method, we first derive the $PH_{norm}$\,distribution d$N$/d$PH_{norm}$ for a normalised current pulse: $I(t') = 1/\tau \cdot e^{-t'/\tau}$ for $t' \geq 0$ and $I(t') = 0 $ for $t' < 0$.
% The pulse height for a Geiger discharge  at the time $t$ relative to the start of the gate is given by Eq.\,\ref{equ:PHt} with $Q_0 = 1$:

  \begin{equation}\label{equ:PHn}
    PH_{norm}(t) = \left\{
           \begin{array}{ll}
           e^{\frac{t} {\tau}} \, \big(1 - e^{-\frac{t_{gate}} {\tau}}\big)
             & \hbox{\rm{for}\,\, $t < 0$,} \\
    \vspace{1mm}
             1 - e^{-\frac{t_{gate} - t} {\tau}}
             & \hbox{\rm{for}\,\, $ 0 \leq t < t_{gate}, $} \\
             0 & \hbox{\rm{for}\,\, $ t \geq t_{gate}$.}
           \end{array}
         \right.
  \end{equation}

 Next the time $t_0 (< 0)$ is introduced.
 $DCR$\,pulses with $t \leq t_0$ are assigned to the pedestal, and  $DCR$\,pulses with $t_0 < t \leq 0$ result in $PH_{norm}$\,values in the range
% The $PH_{norm}$ range for $DCR$\,\,pulses which occur before the start of the gate in the time interval between $t = t_0 < 0$ and $t = 0$ is:
 $PH_{norm}^{min}(t_0) = e^{t_0 /\tau}\, (1 - e^{-t_{gate}/\tau}) $ and  $PH_{norm}^{max} = (1 - e^{-t_{gate}/\tau} )$.
 The $PH_{norm}$\,distribution is given by $\mathrm{d}N/\mathrm{d}PH_{norm} = (\mathrm{d}N/\mathrm{d}t)/|\mathrm{d}PH_{norm}/\mathrm{d}t| = DCR / |\mathrm{d}PH_{norm}/\mathrm{d}t|$ with $PH_{norm}(t)$ given by Eq.\,\ref{equ:PHn} and the dark-count rate  $\mathrm{d}N/\mathrm{d}t \equiv DCR$.
 The result is: d$N$/d$PH_{norm} = (DCR \cdot \tau)/PH_{norm}$.
 So far, correlated noise pulses, which shift entries from the $npe = 1$\,peak to higher values, are not taken into account.
 This will be discussed below.
 A similar derivation for \textit{DCR} pulses occurring during the gate, $0 \leq t \leq t_{gate}$,
 shows that the $PH_{norm}$\,values are in the range 0 to $PH_{norm}^{max}$ with the distribution d$N$/d$PH_{norm} = (DCR\cdot \, \tau)/(1 - PH_{norm})$.

 Thus the $PH_{norm}$ distribution for random $DCR$ events with unit charge for $t_0 \leq t \leq t_{gate}$
 is
 \begin{equation}\label{equ:Eq3}
    \frac{\mathrm{d} N} {\mathrm{d} PH_{norm}} = (DCR \cdot \, \tau) \cdot \Big(\frac{1} {PH_{norm}} + \frac{1} {1 - PH_{norm}} \Big),
 \end{equation}
 where the first term has to be evaluated for $PH_{norm}$\,values between $PH_{norm}^{min}(t_0)$ and $PH_{norm}^{max}$, and the second term between 0 and $PH_{norm}^{max}$.
  For \textit{DCR} pulses outside of the time interval $ t_0 < t \leq t _{gate}$,  $PH_{norm}$ is set to zero, and their contribution to the $PH_{norm}$ spectrum is:
  \begin{equation}\label{equ:Eq4}
   \frac{\mathrm{d}\textit{N}} {\mathrm{d}PH_{norm}} = \big(1 - DCR \cdot (t_0 + t_{gate})\big) \cdot \delta (PH_{norm} = 0),
 \end{equation}
 where $\delta(x)$ is the Dirac delta function.

 The choice of the value for $t_0$ is not critical, as long as $PH_{norm}^{min}(t_0)$ is small compared to the width of the pedestal distribution.
 Increasing the value of $|t_0|$ only shifts events from Eq.\,{\ref{equ:Eq4} to Eq.\,{\ref{equ:Eq3} in the region of $PH_{norm}^{min}(t_0)$.
 For the fits discussed below, $t_0 = - a \cdot \tau$ with $a = 5$ has been used, and it has been verified that varying $a$ between 3.5 and 7.5 does not affect the results.

 Using a binned maximum-likelihood method, the measured $PH$\,spectra in the region of the $npe=0$ and the $npe =1$ peaks are fitted by the sum of  Eqs.\,\ref{equ:Eq3} and \ref{equ:Eq4} with $PH_{norm}$ scaled by $Gain =  PH_1 - PH_0$ and shifted by $PH_0$, convolved with a Gaussian of variance $\sigma ^2$, and normalised to the number of measured events.
 As the function described by Eq.\,\ref{equ:Eq3} is non continuous with sharp peaks at $PH_{norm}^{min}$ and $PH_{norm}^{max}$, which depend on the free parameter \textit{DCR}, special care has to be taken for the convolution:
 Eq.\,\ref{equ:Eq3} is evaluated in bins of constant width in $\ln (PH_{norm})$ for $PH_{norm} < 0.5$, and in bins of constant width in $\ln (1-PH_{norm})$ for $PH_{norm} \geq 0.5$, and then convolved with Gaussians in the range $\pm \,5 \, \sigma $.
 The free parameters of the fit are:
 the position of the $npe = 0$ and $npe = 1$ peaks, $PH_0$ and $PH_1$, the dark-count rate, $DCR$, the noise term, $\sigma $, and the normalisation.

 The results of the fits for 29.5, 31.0, 33.0, and 35.0\,V are shown as dotted lines, labeled "Fit\,1", in Fig.\,\ref{fig:FigDCR}.
 The data, in particular also the region in-between the $npe=0$ and the $npe=1$ peaks, which is populated by events with only partial overlap of the dark-count  pulses with the gate, are well described.
 The number of entries between the $npe =0$ and the $npe=1$ peak is sensitive to the pulse decay time, $\tau $.
 The value $\tau = 19.95$\,ns, obtained from the delay curve in Sect.\,\ref{sect:Delay}, provides a good description of the data.
 Changes by $\pm \, 2.5$\,ns spoil this agreement.
 Deviations between model and fit are observed at the low $PH$ tail of the $npe=0$ peak at 35\,V.
  At this voltage the \textit{DCR} is highest and the AC coupling produces deviations from the Gauss distribution of the pedestal peak.
 As the model considers neither multiple dark-count pulses nor correlated noise, the function drops to zero above the $npe =1 $ peak.
 In addition, the value of DCR obtained from "Fit1" will be incorrect, if the probability for correlated noise is high.

 Multiple dark-count  pulses as well as prompt cross-talk and after-pulses from the dark-count  pulses, result in \textit{PH} values beyond the $npe = 1$ peak.
 The following assumptions are made to derive a model, which considers \textit{PH} values up to $npe = 4$:
 Poisson statistics with a mean of $DCR \cdot (|t_0| + t_{gate})$ for the probability distribution of the dark-count  pulses, and for each dark-count pulse a Borel branching process for the probability distribution of additional prompt cross-talk pulses.
 For further details we refer to Appendix\,B.
%  and  multiplication of these probabilities with the convolutions of the corresponding \textit{PH} distributions.
 For two reasons, after-pulses are not taken into account:
 As pointed out in Sect.\,\ref{sect:PHLED}, the time dependence of the after-pulses is not known, and the additional convolutions with delayed after-pulses significantly complicate the already complex model.

 The results of binned maximum likelihood fits of the model to the data of 29.5, 31.0, 33.0, and 35.0\,V are shown in Fig.\,\ref{fig:FigDCR} as solid lines, labeled "Fit\,2".
 The model qualitatively describes the data also for \textit{npe}\,>\,1.
 However, in particular at higher voltages, there are significant discrepancies for \textit{PH} values above the $npe = 1$ peak:
 More events than predicted are observed in-between the $npe$ peaks, and the peaks in the data are less pronounced.
 We ascribe this to the neglect of the after-pulses.

 Next the fit results are compared to the results of a method, which is similar to the standard method of determining the $DCR$.
 The $npe = 0$ peak is  fitted by a Gaussian in order to determine its mean value, $PH_0$, and its variance, $\sigma ^2$.
 For $Gain$ the value determined from the $PH$\,measurements with the SiPM illuminated by a pulsed LED, presented in Sect.\,\ref{sect:PHLED}, is used.
 The value of $f_{0.5}$ is obtained from the fraction of entries above $PH_0 + 0.5 \cdot Gain$, and for $f_{0.5}^{corr}$ the fraction of entries in the tail of the Gauss function fitted to the $npe = 0$ peak is subtracted from $f_{0.5}$.
 The dark-count rate is calculated using $DCR = f_{0.5}^{corr} / t _{gate}$, where $t _{gate} = 100$\,ns is used.
%  compatible with the value obtained from the fits to the delay curve discussed in  Sect.\,\ref{sect:Delay}.

 In Fig.\,\ref{fig:FigDCRXT}a the \textit{DCR} results for the two methods are shown as a function of voltage.
 The $DCR$ increases approximately linearly with voltage and reaches a value of about 220\,kHz at 35\,V, which is about 8\,V above the breakdown voltage.
 At 29.5\,V, where the $npe =0$ and $npe =1$ distributions overlap, the correction to $f_{0.5}$ is required.
 Above 29.5\,V the $f_{0.5}$ and the $f_{0.5}^{corr}$ results are identical.
 The results of the \textit{DCR} fit and the $f_{0.5}$ method agree up to a voltage of 32\,V.
 For higher voltages the \textit{DCR} fit results are somewhat lower, with a maximal difference of 7\,\% at 35\,V.
 We assume that the reason for this discrepancy is the neglect of after-pulses in the \textit{DCR} model, and consider the $f_{0.5}^{corr}$ results to be more reliable.

    \begin{figure}[!ht]
%%D:\sync\Junk\DetectorWork\SiPM\Ketek\NIrradiation\Analysis\MP15\QDC\QDC-150827\Dark-Fit-03.xmcd
   \centering
   \begin{subfigure}[a]{0.5\textwidth}
   \centering
    \includegraphics[width=\textwidth]{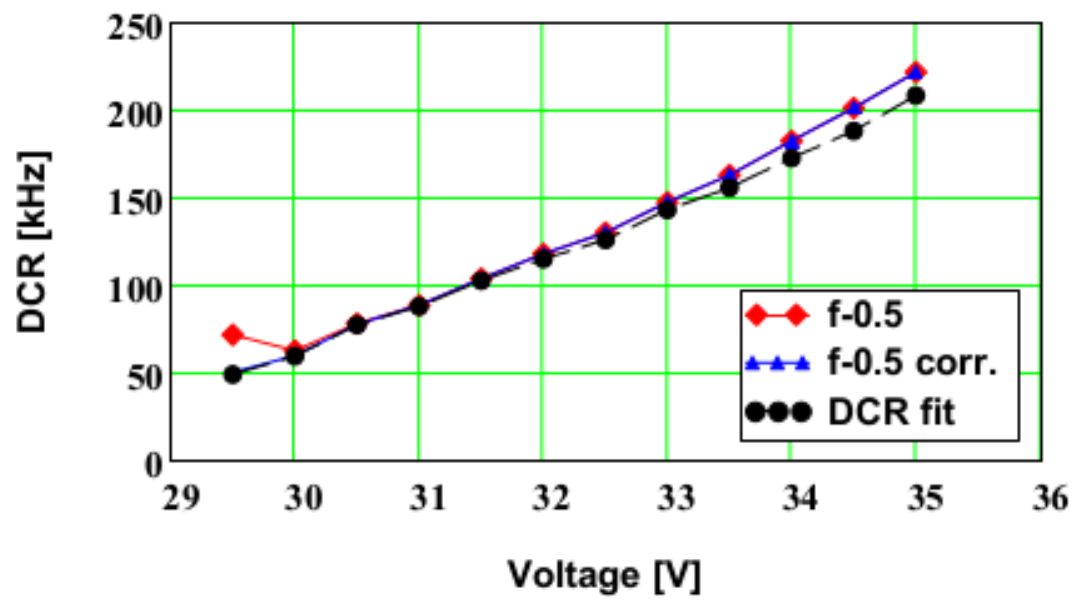}
    \caption{ }
   \end{subfigure}%
    ~
   \begin{subfigure}[a]{0.5\textwidth}
    \includegraphics[width=\textwidth]{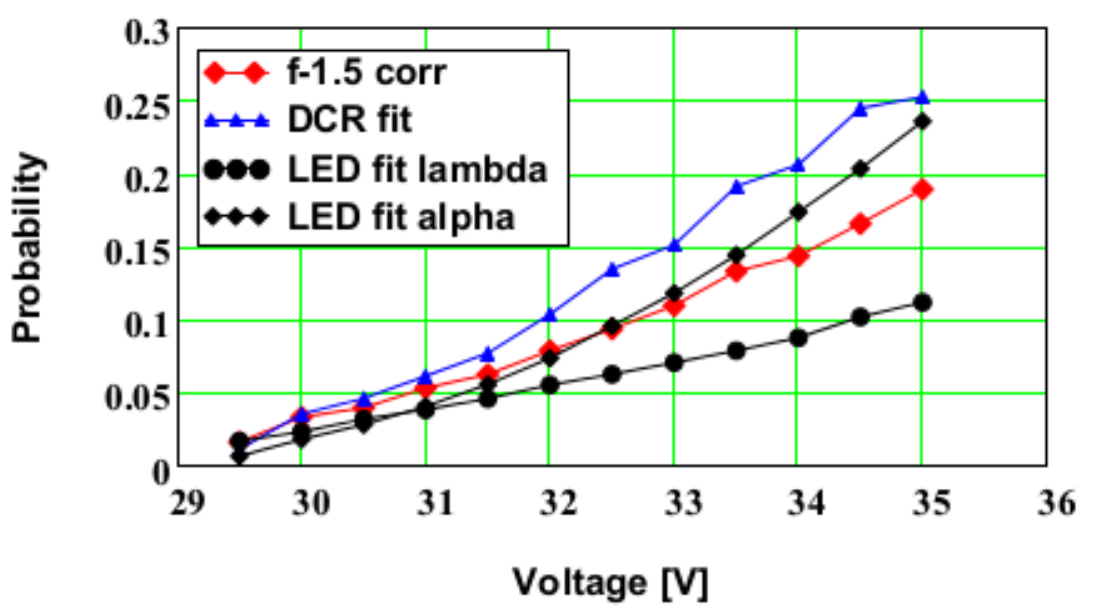}
    \caption{ }
   \end{subfigure}%
   \caption{ Comparison of the results of different methods of determining the voltage dependence of
    (a) the dark-count-rate, \textit{DCR}, and
    (b) the cross-talk and after-pulse probabilities.
   }
  \label{fig:FigDCRXT}
 \end{figure}

 The correlated noise, \textit{CN}, the combined effect of prompt cross-talk and after-pulsing, is determined using $CN = f_{1.5}/f_{0.5}^{corr}$, where $f_{1.5}$ is the fraction of entries in the \textit{DCR} spectrum above $PH_0 + 1.5\, Gain$.
 The results, labeled "f-1.5 corr", are shown in Fig.\,\ref{fig:FigDCRXT}b and compared to the results of the \textit{DCR} fit, labeled "DCR fit".
 The \textit{DCR} fit results are consistently higher, with a difference, which increases with voltage.
 Also shown are the results for the prompt cross-talk, labeled "LED fit lambda", and for the after-pulses, "LED fit alpha", from the fit to the LED data presented in Sect.\,\ref{sect:PHLED}.
 Shown are $1 - e^{- \lambda }$ and $\alpha $, which correspond to the probability that a Geiger discharge causes one or more prompt discharges or an after-pulse, respectively.
 We find it difficult to interpret the differences, as they correspond to different quantities.
 However, as the \textit{DCR} value from the fit is somewhat smaller than from the $f_{0.5}$ method, and  the mean \textit{PH} of the after-pulses is significantly smaller than for a single Geiger discharge, there may  not be a real discrepancy.
%  for the correlated noise between the results from the \textit{LED} fits and the $f_{1.5}$ method.

 The \textit{DCR} fits also determine \textit{Gain}.
 In Fig.\,\ref{fig:FigResults}b the results are compared to the values from the low-light fit.
 Typical differences are  1 to 2\,\%, with a maximal deviation of 2.9\,\%.

 \emph{To summarise}:
 A model has been developed, which attempts to describe the dark-count pulse-height spectra of SiPMs.
 It takes into account the random arrival times of dark pulses, multiple dark pulses and prompt cross-talk, but neglects after-pulses.
 The model is fitted to the data to determine the dark count rate, \textit{DCR}, the gain, \textit{Gain}, and the correlated noise, \textit{CN}.
 A qualitative agreement with the measured spectra is achieved, however at higher voltages the region above the $npe = 1$ peak is only approximately described.
 The differences are ascribed to the neglect of after-pulses.
 The comparison with the results of the standard method shows satisfactory agreement for \textit{DCR}.
 For the correlated noise, differences of up to 25\,\% are found.
 We suspect that the cause is the neglect of the after-pulses, and consider the result of standard method more trustworthy.
 The values of \textit{Gain} determined agree within <\,3\,\% with the values determined from fits to the low-light spectra.

  \section{Application of the results}
  \label{sect:Application}
% \subsection{Current-breakdown voltage $V_I$}
% \label{subsect:Current}

 \subsection{Resolution of photon detection and excess noise factor}
  \label{sect:ENF}

 The question addressed in this section is:
 At which voltage is the best resolution for the measurement of the number of photons from a light source achieved?
 This is e.\,g. relevant for the energy measurement using SiPMs coupled to scintillators.
 As a function of voltage, the photon-detection efficiency increases, but the excess noise from prompt cross-talk and after-pulsing also increases.
 Thus there will be an optimum.
 The relevant quantity for a given light source is the resolution, $Res = \sqrt{var} / mean$, where $mean$ is the mean and $var$ the variance of the  \textit{PH} distribution after subtracting the pedestal value.
 The dependence of \textit{Res} on voltage for the low-light data is shown in Fig.\ref{fig:ENF}\,a.
 In addition to $Res$ calculated from the moments of the measured \textit{PH} spectra, labeled "Data", and from the fit curve, "Fit", the values for the Poisson distribution $Res_{Poisson} = 1/\sqrt{\mu }$ and for the Generalised Poisson distribution $Res_{Gen.\,Poisson} = 1/\sqrt{\mu \,(1 - \lambda)}$, with the values determined in Sect.\,\ref{sect:PHLED}, are shown.
 Fig.\ref{fig:ENF}\,a shows that between 34 and 35\,V the improvement in resolution due to the relative increase of the photon-detection efficiency by 6.3\,\% is largely compensated by the increase in excess noise:
 $Res$ improves by 0.7\,\% only.
 It has been checked that the same voltage dependence of \textit{Res} is obtained from the high-light data.
 Thus the optimal voltage is about 34\,V.
  At higher voltages the resolution does not improve anymore, however, excess noise and dark-count rate increase significantly.

% Mathcad: Junk/DetectorWork/SiPM/Ketel/N-Irradiation/Analysis/MP15/QDC/QDC-Results02.xmcd
   \begin{figure}[!ht]
   \centering
   \begin{subfigure}[a]{0.5\textwidth}
   \centering
    \includegraphics[width=\textwidth]{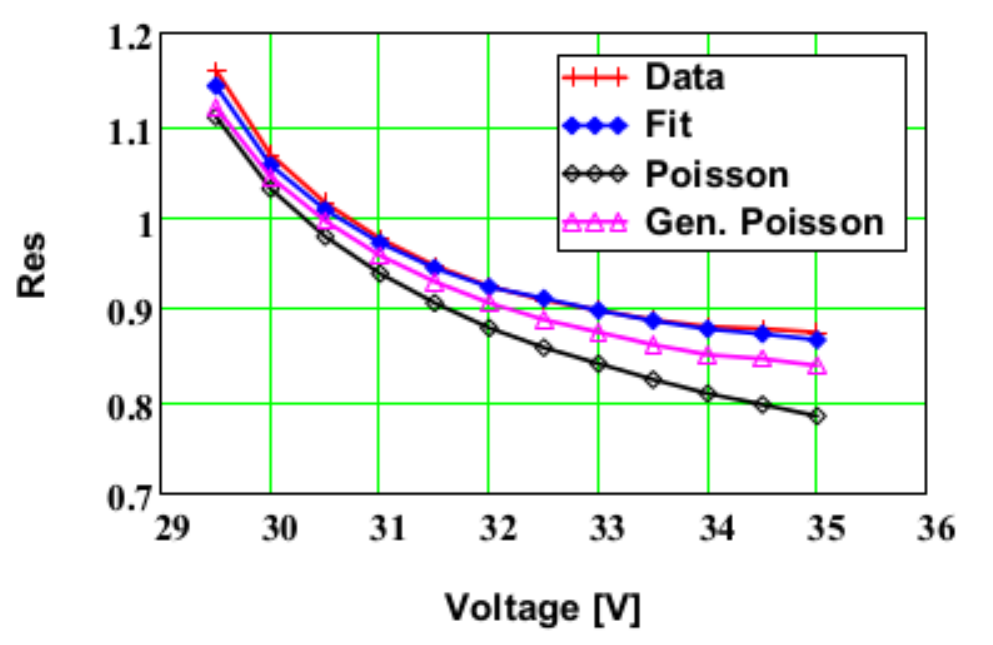}
    \caption{ }
   \end{subfigure}%
    ~
   \begin{subfigure}[a]{0.5\textwidth}
    \includegraphics[width=\textwidth]{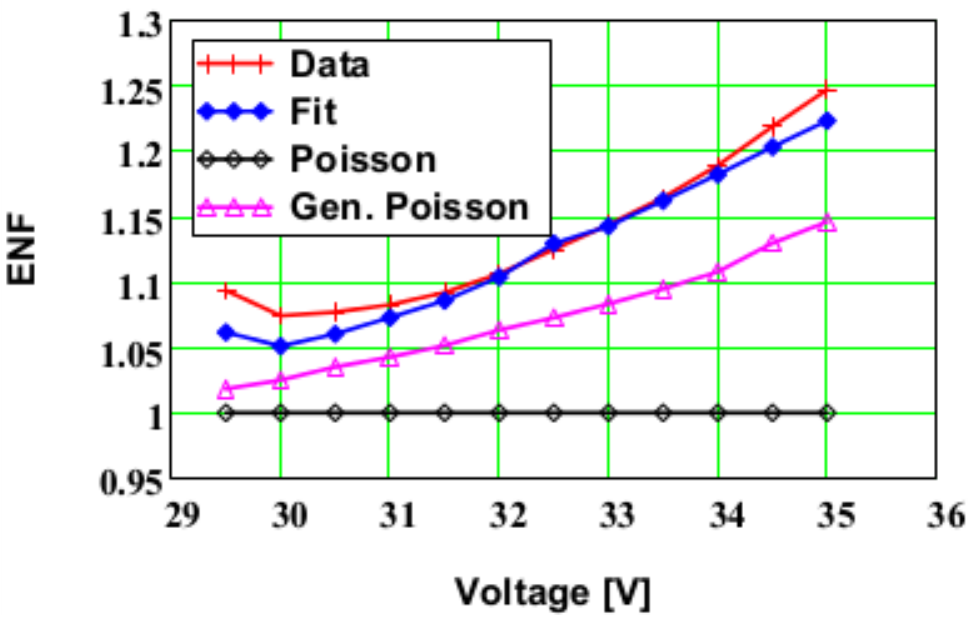}
    \caption{ }
   \end{subfigure}%
   \caption{ (a) Resolution, $Res = \sqrt{var}/mean $, for the pulse-height spectra of the low-light measurements, and (b) excess noise factor, \textit{ENF}, for
    "Poisson", the Poisson distribution,
    "Gen.\,Poisson", the Generalised Poisson distribution with the parameters $\mu $ and $\lambda$ obtained in Sect.\,\ref{sect:PHLED},
    "Fit", the fit function described in Sect.\,\ref{sect:PHLED}, and
    for "Data", calculated from the moments \textit{mean} and \textit{var}  of the measured spectra.
     }
  \label{fig:ENF}
 \end{figure}

 To quantify the worsening of the resolution by excess noise, the excess-noise-factor, \textit{ENF}, is commonly used.
 For a distribution $i$, $ENF_i$ is defined as
 \begin{equation}\label{equ:ENF}
   ENF_i = \big(var_i / mean_i^2 \big) / \big(var_{Poisson} / mean_{Poisson}^2 \big) = \mu \cdot (var_i / mean_i^2 ).
 \end{equation}
 Fig.\ref{fig:ENF}b shows  \textit{ENF}$_i$ of the low-light data for the Generalised Poisson distribution, "Gen.\,Pois\-son", for the fit function, "Fit", and for the measured spectra, "Data".
 The nonphysical increase of \textit{ENF} for "Fit" and "Data" between 30 and 29.5\,V is caused by the contribution of the pedestal width to \textit{var}, which can be corrected by subtracting $\sigma_0 ^2$.

 Eq.\,\ref{equ:ENF} also provides a simple way to determine $ENF$ using low-light spectra:
 Determine from the fraction of entries in the pedestal peak corrected for dark pulses, $f_0$, calculate $\mu = -\ln(f_0)$ using Poisson statistics, evaluate $mean$ and $var$ from the spectrum, and use Eq.\,\ref{equ:ENF} to obtain \textit{ENF}.
 The method has the advantage that it is simple to use and that no fits of pulse-height spectra are required.
 In Sect.\,\ref{sect:NPE} \textit{ENF} is used to determine the average number of photons initiating a Geiger discharge and the SiPM gain from the moments \textit{mean} and \textit{var} of the pulse-height distribution measured with a pulsed light source.
  This method also works, when the individual \textit{npe} peaks are not separated, and the standard methods of gain determination cannot be applied.

 \subsection{Determination of the number of detected photons and of the gain}
  \label{sect:NPE}

   For calibrating and monitoring a detector with many SiPMs, a simple and robust method to determine the mean number of detected photons, $\langle N_\gamma \rangle $, and the overall gain of the set-up is highly desirable.
   If the \textit{ENF} values of the SiPMs are known, and the \textit{PH} spectra from a pulsed light source are recorded in-situ, inverting Eq.\,\ref{equ:ENF} gives
  \begin{equation}\label{equ:Ngamma}
   \langle N_\gamma \rangle \equiv \mu =  ENF \Big( \frac {mean ^2} {var} \Big).
  \end{equation}
   The \textit{ENF} values of the SiPMs can be determined before their installation into the detector with the method described in Sect.\,\ref{sect:ENF} using low-light \textit{PH}-spectra.
  The proposed method also works, if the different \textit{npe} peaks can not be separated, either because of a high number of photons, a coarse binning of the \textit{PH} spectra, high dark-count rates e.g.\,due to radiation damage or operation at high temperature, or because of electronics noise.
  In the case of a significant noise, the variance of the pedestal distribution has to be subtracted from the variance of the spectrum recorded with light.
  We note that for a high number of photons, the Generalised Poisson distribution approaches a Gauss distribution, and a fit to the \textit{PH} spectrum by a Gauss function can also be used for determining the mean and the variance.

  In Fig.\,\ref{fig:MuHi}a the value of $\mu$ determined in Sect.\,\ref{sect:PHLED}, denoted "measured", is compared to the value calculated using Eq.\,\ref{equ:Ngamma}, denoted "calculated", for the high-light data.
  Here $\sigma_0 ^2$ has been subtracted from $var$.
  The agreement is very good, demonstrating the validity of the method: The difference is 4\,\% at 29.5\,V, and below 1\,\% above 30.5\,V.
  The method described above is routinely used for the calibration of the MAGIC telescope\,\cite{Gaug:2005}.

   \begin{figure}[!ht]
   \centering
   \begin{subfigure}[a]{0.5\textwidth}
   \centering
    \includegraphics[width=\textwidth]{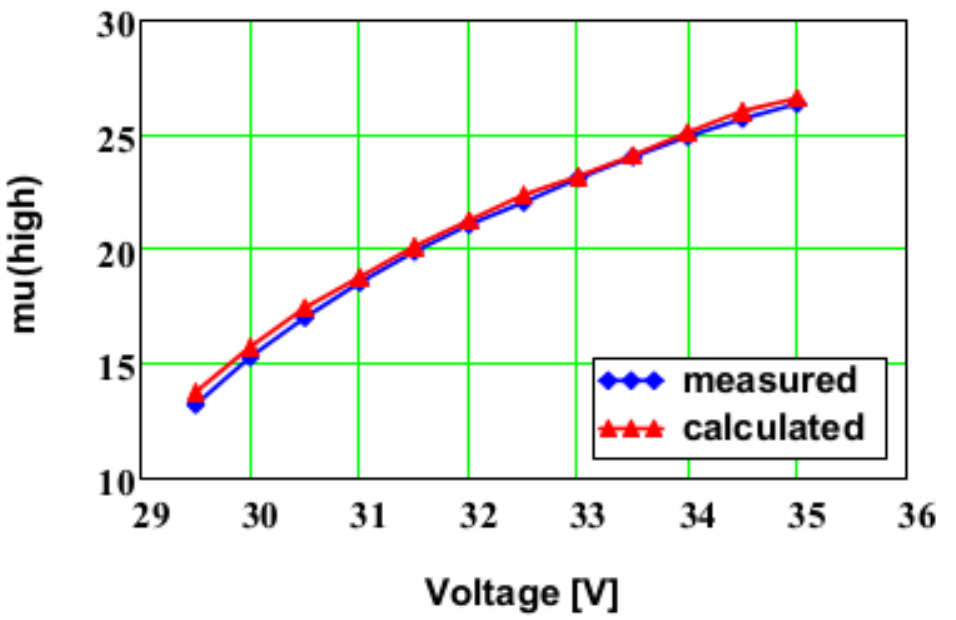}
    \caption{ }
   \end{subfigure}%
    ~
   \begin{subfigure}[a]{0.5\textwidth}
    \includegraphics[width=\textwidth]{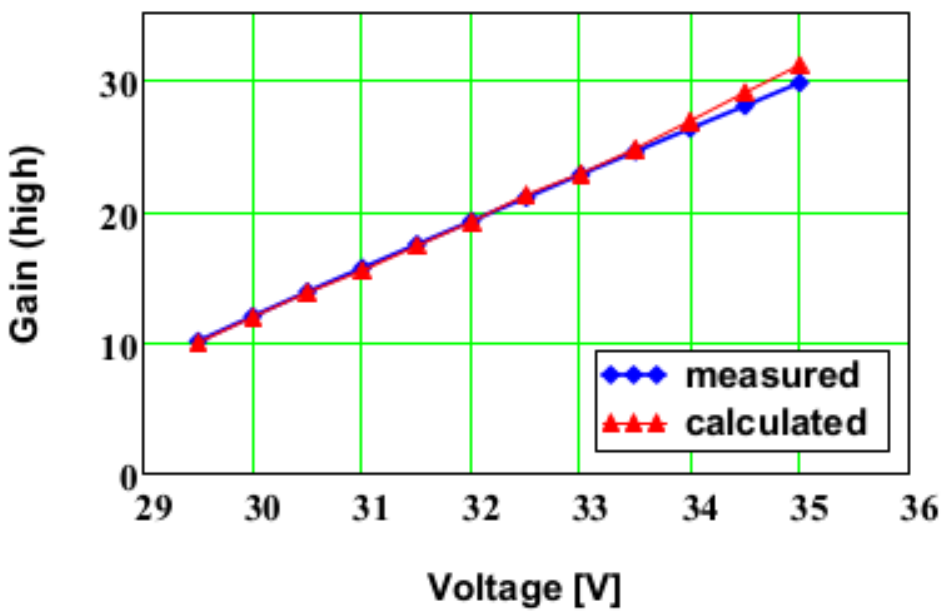}
    \caption{ }
   \end{subfigure}%
   \caption{ Comparison of (a) the mean number of detected photons, and (b) the gain for the high-light data using the fits (measured) and the method of the first and second moments of the measured pulse-height distributions, described in the text (calculated).
     }
  \label{fig:MuHi}
 \end{figure}

 In a similar way, the combined gain of the SiPM and the readout, $Gain'$, can be obtained, if the statistics of the combined effect of prompt cross-talk and after-pulsing is similar to the statistics of the Generalised Poisson distribution, \textit{GP}.
 This is the case if the after-pulse probability $\alpha $ does not exceed $\approx 25$\,\%.
 The relation used is:
   \begin{equation}\label{equ:Gain}
   Gain' = \frac{1} {ENF^2} \Big( \frac{var} {mean} \Big).
  \end{equation}
 It can be derived using the following properties of the Generalised Poisson distribution, \textit{GP}:
 $mean _{GP} = \mu /(1 - \lambda)$ and
 $var _{GP} = \mu /(1 - \lambda)^3.$
 Using Eq.\,\ref{equ:ENF} we find $ENF_{GP} = 1/(1-\lambda)$.
 For the measured \textit{PH} distribution with the gain,
 $Gain'$, $var = Gain'^{\,2} \cdot \mu/(1 - \lambda)^3 = Gain'^{\,2} \cdot \mu \cdot ENF^{\,3}$, and
 $mean = Gain' \cdot \mu/(1 - \lambda) = Gain' \cdot \mu \cdot ENF$,
 and the ratio results in Eq.\,\ref{equ:Gain}.

 In Fig.\,\ref{fig:MuHi}\,b the gain values determined directly from the spectra, labeled "measured", are compared to the values "calculated" using Eq.\,\ref{equ:Gain}.
 Up to a voltage of 33.5\,V both values agree to within 1\,\%.
 For higher voltages, where the probability of after-pulses, $\alpha $, is higher than the probability of prompt cross-talk, $\lambda $, as shown in Fig.\,\ref{fig:FigResults}\,d, the gain values are overestimated.
 The maximum deviation is 4.5\,\% at 35\,V.
 As can be seen from Fig.\,\ref{fig:Fig8}d, for the high-light data at 35\,V, the individual \textit{npe} peaks are not separated, and neither $\langle N_\gamma \rangle$ nor $Gain'$ could have been determined using the standard methods.

 \emph{To summarise:}
 Using the method to determine the excess-noise-factor, \textit{ENF}, described in Sect.\,\ref{sect:ENF}, the combined gain of the SiPM and the readout, and the mean number of detected photons, $\langle N_\gamma \rangle $, can be obtained from the mean and the variance of the measured pulse-height spectra.
 The method is simple and also applicable for high $\langle N_\gamma \rangle $ values, as long as saturation effects due to the finite number of pixels can be ignored.
 Taking saturation effect into account, appears not to be too complicated.
% One method is discussed in Ref.\,\cite{Xu:Thesis}.
% For $ENF = 1$, which corresponds to Poisson statistics, the formulae agree with the ones frequently used for photo-detectors with a small excess noise, like vacuum photomultipliers.

% \subsection{Photon-detection threshold for high dark-count rates}
%  \label{sect:Threshold}

 \section{Conclusions and outlook}
 \label{sect:Conclusions}

 For a KETEK SiPM with 4384 pixels of $15\, \upmu $m$\times 15\, \upmu $m pitch, the pulse-height spectra have been measured for voltages between 2.5 and 8\,V above the break-down voltage at 20$^{\circ}$C for two different intensities of pulsed light and without illumination.
 A model for analysing the pulse-height spectra measured with pulsed light has been developed, which includes the statistics of the photons triggering Geiger discharges, the statistics of prompt cross-talk, and the pulse-height distribution and statistics of after-pulses.
 The model describes the measured pulse-height spectra, including the "background".
%  in-between the peaks from different number of Geiger discharges.
 As far as we know, such a description has not yet been achieved so far.
 From the agreement of the model with the data it is concluded:
 The statistics of the cross-talk from a primary Geiger discharge can be described by a Borel distribution, which results in a Generalised Poisson distribution for the combined statistics of primary Geiger discharges and cross-talk. The statistics of after-pulses from a Geiger discharge can be described by a binomial distribution.
 The pulse-height distribution of the after-pulses has been derived from the data.

 By fitting the model to the pulse-height spectra, the voltage dependence of the SiPM gain, the number of photons initiating Geiger discharges, the cross-talk and the after-pulse probability were obtained.
 The results were used to determine the excess-noise factor due to prompt cross-talk and after-pulses, and the voltage for optimal light-yield resolution.

 Based on the detailed understanding of the SiPM statistics, a  method for calibrating and monitoring the number of photons initiating Geiger discharges and the combined gain of the system SiPM and readout is demonstrated, which appears suitable for detectors with a large number of SiPMs, like calorimeters.
 It uses the mean and the variance of the pulse-height spectra measured in-situ from a pulsed light source, and the excess noise factor, which can be determined in a straight-forward way as part of the pre-installation quality control of the SiPMs.
 For the SiPMs investigated, the accuracy of the gain and the number of photons determined agrees to better than 5\,\% with the fit results, when the light intensity is changed by a factor of 16.

 A model for the pulse-height spectra of dark counts is developed, which takes into account the random occurrence of dark pulses and prompt cross-talk, but neglects after-pulses.
 It provides a good description of the peaks corresponding to zero and one Geiger discharges, and the region in-between, and allows to determine the gain, the dark-count rate and the cross-talk probability.
 The spectra above the one Geiger-discharge peak are only qualitatively described, indicating that also the effect of after-pulses has to be implemented in the model.
 It is shown that the SiPM gain can be determined from pulse-height spectra measured in the dark, without the need of a pulsed light source, which could be of interest for detectors with a large number of SiPMs.

 \section*{Acknowledgement}
 \label{sect:Acknowledgement}
 We would like to thank Florian Wiest and his colleagues from KETEK for providing the SiPMs samples and for fruitful discussions.
 We are also thankful to Peter Buhmann and Michael Matysek for keeping the measurement infrastructure of the Hamburg Detector Laboratory, where the measurements have been performed, in excellent operating conditions.

\input{bibliography-R1}

 \label{sect:Bibliography}

% \newpage

 \appendix
 \renewcommand*{\thesection}{\Alph{section}}
  \section{Appendix A: Pulse-height distribution for an exponential after-pulse probability}
  \label{sect:AppendixA}

%  \section{Appendix II}

 In Sect.\,\ref{sect:PHLED} we have stated that that the observed exponential \textit{PH} distribution for AP pulses for the KETEK SiPM was not expected.
 In this Appendix we derive the dependence expected for an exponential time dependence of the after-pulse probability.
 To simplify the formulae we assume $Gain = 1$,  define $PH_{AP}$ as the additional \textit{PH} from the after-pulse, measure the time $t$ from the start of the Geiger discharge, which causes the after-pulse and assume that $t = 0$ is also the start of the integration by the QDC.
 For the time dependence of the AP probability density an exponential is assumed:
 \begin{equation}\label{dpdtAP}
   \frac{\mathrm{d}p_{AP}} {\mathrm{d}t} = \frac{1} {\tau _{AP}} e^{-t/\tau _{AP}}.
 \end{equation}
% where $\tau _{AP} $ is the tome constant of the $AP$ decays.
 For $PH_{AP}(t)$, the $PH$ of an after-pulse at time $t$ after the primary pulse
  \begin{equation}\label{equ:PHAP}
   PH_{AP}(t)  \left\{
       \begin{array}{ll}  \vspace{2mm}
        =\big(1 - e^{-t/\tau} \big) \cdot \int\limits_0^{t_{gate} - t} \big(\frac{e^{-t'/\tau}} {\tau} \big) \,\mathrm{d}t'   =  \\
        =1+e^{-t_{gate}/\tau} -2\,e^{-t_{gate}/(2\,\tau)}\,\cosh\big(\frac{t_{gate}/2-t} {\tau}\big)
           \end{array}
         \right.
  \end{equation}
 is assumed.
 The first term of the upper line of Eq.\,\ref{equ:PHAP} describes the reduction of $PH_{AP}$ due to the recharging of the pixel with the time constant $\tau$ after a Geiger discharge, and the second term the fraction of the AP signal integrated during the QDC gate of duration $t_{gate}$.
 The function $PH_{AP}(t)$ is symmetric around $t=t_{gate}/2$, zero at $t=0$ and $t=t_{gate}$, and the value of the maximum at $t=t_{gate}/2$ is:
 \begin{equation}\label{PHAPmax}
   PH_{AP}^{max} = \big( 1 - e^{-t_{gate}/(2\,\tau)}\big)^2.
 \end{equation}
 To derive the probability density distribution $\mathrm{d}p_{AP}/\mathrm{d}PH_{AP}$, we use
 $\frac{\mathrm{d}p_{AP}} {\mathrm{d}PH_{AP}} =
 \frac{\mathrm{d}p_{AP}} {\mathrm{d}t} /
 | \frac{\mathrm{d}PH_{AP}} {\mathrm{d}t} |$ to obtain
  \begin{equation}\label{dpdPHAP}
   \frac{\mathrm{d}p_{AP}} {\mathrm{d}PH_{AP}} = \frac{\tau} {\tau_{AP}} \, \frac{e^{-t(PH_{AP})/\tau_{AP}} } {2\, e^{-t_{gate}/(2\,\tau)} |\sinh\big(\frac{t_{gate}/2-t(PH_{AP})} {\tau} \big)|},
 \end{equation}
 which is valid in the $PH_{AP}$ range between 0 and $PH_{AP}^{max}$.
  The function $t(PH_{AP})$ is obtained by inverting  Eq.\,\ref{equ:PHAP}.
 For more than one AP, the probability distribution has to be convolved with itself, and finally the complete probability density is convolved by a Gaussian to account for noise.
 The convolutions are performed using the Fast-Fourier-Transform.
 We note that in this model, for certain values of the ratio $t_{gate}/\tau$, peaks, caused by after-pulses, appear at $PH_{AP}^{max}$ in-between the $npe$ peaks.
 Such intermediate peaks have actually been observed for a MPPC from Hamamatsu.

   \section{Appendix B: Fit function for the DCR spectra }
  \label{sect:AppendixB}

 In this appendix we sketch the derivation of the formulae used in Sect.\,\ref{sect:PHdark} to fit the $PH_{dark}$ spectra including the  $npe = 4$  peak.
 To simplify the formulae, we use the normalised pulse height, $x$, with $x = 0$  the mean of the pedestal and $x = 1$ the mean of the $npe = 1$ peak, and derive the formulae without smearing due to electronic noise and gain differences between  pixels.

 We call $f^{(1)}(x)$ the normalised $PH$\,probability distribution derived from Eq.\,\ref{equ:Eq3}.
 For \emph{i} $DCR$\ pulses in the time interval between $t_0$ and $t_{gate}$ and no prompt cross-talk, the normalised $PH$ distribution, $f^{(i)}(x)$, is the $i$-fold convolution of $f^{(1)}(x)$, calculated using the $i$-th power of the Fourier transform of $f^{(1)}(x)$.
 Prompt cross-talk just stretches the $x$ values of the $PH$ probability distribution, and the distribution for the $j$-fold cross-talk of a single $DCR$ pulse is given by $h^{(j)}(x) = f^{(1)}((j+1)\cdot x)/(j+1)$.
 The distributions for arbitrary combinations of $DCR$ and cross-talk pulses are convolutions of $f^{(i)}(x) \otimes h^{(j)}(x)$.
 Examples are shown in  Table\,\ref{tab:Table1}, which presents for the $i, j$ values possible for $npe \le 4$ the combinatoric factors and the corresponding functions.
% The distribution for $i\,DCR$ pulses of which $j \le i$ have a cross-talk pulse, is given by the convolution $f^{(j-1)} \otimes h^{(j)}$.
 The sum of these functions multiplied with the appropriate combinatoric factors, $Comb$,  and the probability products $Poisson \cdot Borel$ gives the normalised $PH$\,spectrum.
% Table\,\ref{tab:Table1} presents for the $i, j$ values possible for $npe \le 4$ the  combinatoric factors and the corresponding functions.
 Finally, after the transformation  $ x \rightarrow PH_0 + Gain \cdot x$ and the convolution with a Gauss function, the resulting function is fitted to the measured $PH$ distributions.

 \begin{table} [!ht]
  \caption{Parameters and functions used for the fit function for the $PH_{dark}$ spectrum;
   $npe$: total number of Geiger discharges including prompt cross-talk in the time interval $t_0$ to $t_{gate}$;
   $P_{i}$: Poisson probability for $i$ primary dark pulses,
   $B_{j}$: Borel probability for $j$ prompt cross-talk pulses,
   $Comb$: number of permutations for the given configuration,
   and Function: $PH$ distribution for the given configuration.
    }
%   { \renewcommand{\arraystretch}{1.1}
  \centering
   \begin{tabular}{c|c|c|c|c}
%  \hline
  $npe$ & $Comb$ & $Poisson_i$ & $Borel_j$ & Function \\
   \hline \hline
  0 & 1 & 0 & $-$ & $\delta (x)$ \\
    \hline
  1 & 1 & $P_0$ & $B_0$ & $f^{(1)}$ \\
    \hline
  2 & 1 & $P_1$  & $B_1$ & $h^{(1)}$ \\
%    \hline
  2 & 1 & $P_2$ & $(B_0)^2$ & $f^{(2)}$ \\
    \hline
  3 & 1 & $P_1$ & $B_2$ & $h^{(2)}$ \\
%    \hline
  3 & 2 & $P_2$ & $B_0\cdot B_1$ & $f^{(1)}\otimes h^{(1)}$ \\
%    \hline
  3 & 1 & $P_3$ & $(B_0)^3$  & $f^{(3)}$ \\
    \hline
  4 & 1 & $P_1$ & $B_3$ & $h^{(3)}$ \\
%    \hline
  4 & 2 & $P_2$ & $B_0 \cdot B_2$ & $f^{(1)}\otimes h^{(2)}$ \\
%    \hline
  4 & 1 & $P_2$ & $(B_1)^2$ & $h^{(1)}\otimes h^{(1)}$ \\
%    \hline
  4 & 3 & $P_3$ & $(B_0)^2 \cdot B_1$ & $f^{(2)}\otimes h^{(1)}$ \\
%    \hline
  4 & 1 & $P_4$ & $(B_0)^4$ & $f^{(4)}$ \\
  \hline \hline
   \end{tabular}
  \label{tab:Table1}
%  }
 \end{table}

%\bibliographystyle{unsrt}
%\addcontentsline{toc}{section}{\refname}

%\bibliography{bib/bib}

%\bibliographystyle{unsrtnat}

\end{document}

%% file: bibliography-R1.tex
\section{List of References}